\pdfoutput=1
\documentclass[final,3p,times]{elsarticle}
\usepackage[latin9]{inputenc}
\usepackage{bm}
\usepackage{amsmath}
\usepackage{amssymb}
\usepackage{graphicx}
\usepackage{esint}
\usepackage{multirow}
\usepackage[subfigure]{graphfig}
\usepackage{rotating}
\usepackage{verbatim}
\usepackage{setspace}
\usepackage{epstopdf}
\makeatletter

\usepackage{tabularx}\usepackage{subfigure}\usepackage{subfigure}\usepackage{color}

\usepackage{setspace}

\usepackage{bm}

\journal{}

\makeatother

\color{black}

\begin{document}

\title{Implementation of BVD (boundary variation diminishing) algorithm in simulations of compressible multiphase flows}

\author[ad1]{Xi Deng}

\author[ad1]{Satoshi Inaba}

\author[ad1]{Bin Xie}

\author[ad2]{Keh-Ming~Shyue}

\author[ad1]{Feng Xiao \corref{cor}}

\address[ad1]{Department of Mechanical Engineering, Tokyo Institute of Technology, \\
	4259 Nagatsuta Midori-ku, Yokohama, 226-8502, Japan.}

\address[ad2]{Department of Mathematics, National Taiwan University, 
	Taipei 106, Taiwan}
\cortext[cor]{Corresponding author: Dr. F. Xiao (Email: xiao@es.titech.ac.jp)}

\begin{abstract}
	We present in this work a new reconstruction scheme, so-called MUSCL-THINC-BVD scheme, to solve the five-equation model for interfacial two phase flows. This scheme employs the traditional shock capturing MUSCL (Monotone Upstream-centered Schemes for Conservation Law) scheme as well as the interface sharpening THINC (Tangent of Hyperbola for INterface Capturing) scheme as two building-blocks of spatial reconstruction using the BVD (boundary variation diminishing) principle that minimizes the variations (jumps) of the reconstructed variables at cell boundaries, and thus effectively reduces the numerical dissipations in numerical solutions. The MUSCL-THINC-BVD scheme is implemented to all state variables and volume fraction, which realizes the consistency among volume fraction and other physical variables. Benchmark tests are carried out to verify the capability of the present method in capturing the material interface as a well-defined sharp jump in volume fraction, as well as significant improvement in solution quality. The proposed scheme is a simple and effective method of practical significance for simulating compressible interfacial  multiphase flows.
\end{abstract}

\maketitle

\section{Introduction}
Compressible multiphase flow is one of active and challenging research areas of great importance in both theoretical studies and industrial applications. For example, shock/interface interactions are thought to be crucial to the instability and evolution of material interfaces that separate different fluids as can be observed in a wide spectral of phenomena\cite{instability}. The material interfaces greatly complicate the physics and make problems formidably difficult for analytical and experimental approaches in general. In many cases, numerical simulation turns out to be the most effective approach to provide quantitative information to elucidate the fundamental mechanisms behind the complex phenomena of multiphase flows. 

In comparison to computation of single phase flow, development of numerical methods for multiphase flow faces more challenging tasks. The major complexity comes from the moving interfaces between different fluids that usually associate with strong discontinuities, singular forces and phase changes. Given the numerical methods developed for multiphase incompressible flow with interfaces having been reached a relatively mature stage, the numerical solvers for compressible interfacial multiphase flow are apparently insufficient. For incompressible multiphase flows with moving interfaces where the density and other physical properties, e.g. viscosity and thermal conductivity, are constant in each fluid, the one-fluid model \cite{trg-book} can be implemented in a straightforward manner with an assumption that the physical fields change monotonically across the interface region. So, provided an indication function which identifies the moving interface, one can uniquely determine the physical property fields for the whole computational domain. Some indication functions, such as volume of fluid (VOF) function \cite{vof1,vof2,vof3} and level set function \cite{ls1,ls2,ls3}, have been proposed and proved to be able to well define the moving interface with compact thickness and geometrical faithfulness if solved by advanced numerical algorithms. However, substantial barrier exists when implementing the one-fluid model to compressible interfacial multiphase flow. 

The new difficulties we face when applying the one-fluid model \footnote{More precisely, it should be called single-state model or single-equivalent-fluid (SEF) model\cite{five}.  We call such model SEF in the present paper.}  to compressible interfacial multiphase flow lie in two aspects: 
\begin{description}
	\item (I) Density and energy in compressible flow have to be computed separately in addition to the indication function, hence special formulations are required to reach a balanced state among all variables for the interface cell where a well-defined interface falls in;
	\item (II) The numerical dissipation in the so-called high-resolution schemes designed for solving single phase compressible flow involving shock waves tends to smear out discontinuities in numerical solutions including the material interfaces, which is fatal to simulations of interfacial multiphase flows even if the schemes can produce acceptable results in single phase cases. 
\end{description}

For issue (I) mentioned above, mixing or averaging models that consist of Euler or Navier-Stokes equations along with interface-indication function equations for each of fluid components have been derived and widely used as efficient approximations to the state of the interface cell where two or more species co-exist.  A simple single-fluid model was reported in \cite{yabe1993,xiao1994,yabe2001} for interfacial multiphase compressible flows using either explicit time marching or semi-implicit pressure-projection solution procedure. The latter results in a unified formulation for solving both compressible and incompressible multiphase flows. As the primitive variables are used in these models, the conservation properties are not guaranteed, and thus might not be suitable for high-Mach flows involving shock waves.  Conservative formulations, which have been well-established for single phase compressible flows with shock waves, however may lead to spurious oscillations in pressure or other thermal fields \cite{larrouturou1991,karni1994}. It was found that special treatments are required in transporting the material interface and mixing/averaging the state variables to find the mixed state of fluids in the interfacial cell that satisfies pressure balance across material interface for multiple polytropic and stiff gases \cite{Abgrall96,Abgrall99,Shyue98, saurel1999, Abgrall01}, and for van der Waals and Mie-Gr\"{u}neisen equations of state (EOS) \cite{ Shyue99,Shyue01}. A more general five-equation model \cite{five} was developed for a wide range fluids. These models apply to multiphase compressible flows with either spread interfaces or sharp interfaces. The five-equation model will be used in the present work as the PDE (partial differential equation) set to solve.  

Provided the SEF models with some desired properties, such as hyperbolicity, conservation and well-balanced mixing closure without spurious oscillations in thermal variables, we can in principle implement numerical methods for single phase compressible flow (e.g. standard shock-capturing schemes) to solve multiphase ones. TVD (Total Variation Diminishing) schemes, such as the MUSCL (Monotone Upstream-centered Schemes for Conservation Law) scheme \cite{Van_Leer}, can resolve discontinuities without numerical oscillations, which is of paramount importance to ensure the physical fields to be bounded and monotonic in the transition region. However, TVD schemes suffer from excessive numerical dissipation, which brings the problem (II) listed above to us. The intrinsic numerical dissipation smears out the flow structures including the discontinuities in mass fraction or volume fraction that are used to represent the material interfaces. Consequently, material interfaces are continuously blurred and spread out, which is not acceptable in many applications, especially for the simulations that need long-term computation. Applying high order schemes like WENO (Weighted Essentially Non-Oscillatory) scheme \cite{Jiang} to solve compressible multiphase flow are found in the literatures \cite{weno1,weno2}. However, implementing high order schemes might generate numerical oscillations for compressible multiphase flow with more complex EOS, such as the Mie-Gr\"{u}neisen equation of state. In \cite{weno1}, to reduce numerical oscillation introduced by high order schemes, the state variables have to be cast into characteristic fields. Although with this effort, stability cannot be guaranteed in the long-term computation even using a forbiddingly small of time step. In a recent work \cite{Gao}, to further reduce the numerical oscillations and to deal with complicated EOS, an approximate intermediate state at each cell edges is obtained in a more careful way to conduct characteristic decomposition. Furthermore, high order monotonicity-preserving scheme \cite{He} was used to ensure volume fraction remain bounded. In general, the implementation of high order shock capturing schemes in compressible multiphase flows will increase complexity of algorithm and may invoke computational instability.

In order to keep material interfaces being a compact thickness during computation, special treatments are required to sharpen or steepen the interfaces. The existing methods for this purpose can be categorized into interface-tracking and interface-capturing. Interface-tracking methods like Arbitrary Lagrangian-Eulerian (ALE) \cite{ALE1,ALE2}, free-Lagrange \cite{free_L}, front-tracking \cite{front} and level set/ghost fluid \cite{level1,level2,level3,level4}  have the distinct advantage of treating interface as a sharp discontinuity. For example, ALE and free-Lagrange methods treat interfaces as boundaries of distorted computational grids, which however makes them complicated and computationally expensive when there are large interface deformations and topological changes. Another drawback is that these methods are typically not numerically conservative at material interfaces, which may lead to a wrong prediction about the position of interfaces and shock waves. 

Interface-capturing methods resolve the interfaces on fixed Eulerian grids, using  special numerical techniques to sharpen the interface from spreading out. For example, in \cite{Shukla1,Shukla2,Garrick} the advection equation of the interface function is treated by artificial compression method. As a post-processing approach, anti-diffusion techniques have been introduced by \cite{kokh-lago:anti-diffusiv-jcp2010} and \cite{so:anti}. Another approach is to reconstruct the volume fraction under the finite volume framework by THINC (Tangent of Hyperbola for INterface Capturing) function \cite{xiao_thinc2}. By virtue of the desirable characteristics of the hyperbolic tangent function in mimicking the jump-like profile of the volume fraction field, the sharp interface can be accurately captured in a simple way \cite{Shyue2014,Gao}. However, unlike incompressible multiphase flow, a common occurrence when applying various interface-sharping methods explicitly on the SEF models is that velocity and pressure oscillations may occur across the interface \cite{Shukla2,Shyue2014,so:anti,Shukla1,Tiwari} due to the inconsistency between the physical variables and the sharpened or compressed volume fraction field. As stated in \cite{Shukla1,Shukla2}, in contrast to incompressible flows where density of fluid is fixed, artificial interface sharpening scheme cannot be applied alone to volume fraction function in compressible multiphase cases. In compressible multiphase flows, neither fluid density nor volume fraction alone is sufficient to determine the interface location and the fluid density. As some remedies, density correction equations are formulated in \cite{Shukla1,Shukla2}. In \cite{Shyue2014}, a homogeneous reconstruction has been proposed where the reconstructed volume fraction is used to extrapolate the remaining conservative variables across the interface to ensure the thermal and mechanical consistency across the isolated material interfaces. In \cite{Tiwari}, a consistent compression method has been discussed to maintain equilibrium across interfaces. 

To alleviate the defect of shock capturing scheme when solving single-fluid model for compressible multiphase flows, a novel spatial reconstruction is presented in this work to resolve contact discontinuities including material interfaces with substantially reduced numerical dissipation, which then maintains the sharpness of the transition layer of material interfaces throughout even long term computations. The scheme, so-called MUSCL-THINC-BVD, implements the boundary variation diminishing (BVD) algorithm \cite{Sun,Xie} with the traditional MUSCL scheme and the interface-sharpening THINC scheme as two building-blocks for reconstruction. The BVD algorithm choose a reconstruction function between MUSCL and THINC, so as to minimize the variations (jumps) of the reconstructed variables at cell boundaries, which in turn effectively removes the numerical dissipations in numerical solutions. More importantly, we apply MUSCL-THINC-BVD scheme to all state variables and volume fraction, so sound consistency is achieved among volume fraction and other physical variables. Resultantly, the manipulations to the physical variables according to the volume fraction in other existing methods are not needed in the present method. The numerical model is formulated under a standard finite volume framework with a Riemann solver in the wave propagation form \cite{wave1}. The numerical results of benchmark tests verify the capability of the present method in capturing the material interface as a well-defined sharp jump in volume fraction, as well as significant improvement in solution quality. 

The format of this paper is outlined as follows.
In Section~\ref{sec:model}, the governing equations of the five-equation model and closure strategies are stated. In Section~\ref{sec:methods},  after a brief review of the finite volume method in wave-propagation form for solving the
quasi-conservative five-equation model, the details of the new MUSCL-THINC-BVD scheme for spatial reconstruction are presented. 
In Section~\ref{sec:results}, numerical results of benchmark tests are presented in comparison with other high-order methods. Some concluding remarks end the paper in Section~\ref{sec:conclusion}.

\section{Computational models \label{sec:model}}

\subsection{Governing equations}
In this work, the inviscid compressible two-component flows are formulated by the five-equation model developed in \cite{five}. By assuming that the material interface is in mechanical equilibrium of mixed pressure and velocity, the five-equation model consists of two continuity equations for  phasic mass, a momentum equation, an energy equation and an advection equation of volume fraction as follows
\begin{equation}
\label{eq:5eqns} 
\begin{aligned}
&\frac{\partial}{\partial t} \left ( \alpha_{1} \rho_{1} \right ) + 
\nabla \cdot \left ( \alpha_{1} \rho_{1} \mathbf{u} \right )   = 0, \\
&\frac{\partial}{\partial t}  \left ( \alpha_{2} \rho_{2} \right ) + 
\nabla \cdot \left ( \alpha_{2} \rho_{2} \mathbf{u} \right )   = 0, \\ 
&\frac{\partial}{\partial t} \left ( \rho \mathbf{u} \right ) + 
\nabla \cdot \left ( \rho \mathbf{u} \otimes \mathbf{u} \right ) +
\nabla p  = 0, \\
&\frac{\partial E}{\partial t}  + 
\nabla \cdot \left ( E \mathbf{u} +  p \mathbf{u} \right )  = 0, \\
&\frac{\partial \alpha_{1}}{\partial t} + 
\mathbf{u} \cdot \nabla \alpha_{1}  =  0,
\end{aligned}
\end{equation}
where $\rho_{k}$ and $\alpha_{k} \in [0,1]$ 
denote in turn the $k$th phasic density and volume fraction
for $k=1,2$, $\mathbf{u}$ the vector of particle velocity, $p$ the mixture pressure and $E$ the total energy. When considering more than two-phases, the five-equation model can be extended by supplementing additional continuity equations and volume fraction advection equations for each new phase. 

\subsection{Closures strategy}
To close the system, the fluid of each phase is assumed to satisfy the Mie-Gr\"{u}neisen equation of state, 
\begin{equation}
\label{eq:mg-eos}
p_{k} \left (\rho_{k},e_{k} \right )
= p_{\infty,k}(\rho_{k}) + \rho_{k} \Gamma_{k}(\rho_{k}) 
\left ( e_{k}- e_{\infty,k}(\rho_{k}) \right ),
\end{equation}
where $\Gamma_{k} = (1/\rho_{k})(\partial p_{k}/\partial e_{k})|_{\rho_{k}}$ is
the Gr\"{u}neisen coefficient, and $p_{\infty,k}$, $e_{\infty,k}$ are the properly chosen states of the pressure and internal energy along some reference curves (e.g., along an isentrope or other empirically fitting curves) in order to match the experimental data of the material examined \cite{marsh:eos}. Usually, parameters 
$\Gamma_{k}$, $p_{\infty,k}$ and $e_{\infty,k}$ can be taken as functions only of the density. This equation of state can be employed to approximate a wide variety of materials including some gaesous or solid explosives and solid metals under high pressure.

The further closure of the system is completed by defining the mixed volume fraction, density and  internal energy as 
\begin{equation}
\label{eq:mixtures} 
\begin{aligned}
&\alpha_{1}+\alpha_{2}=1,\\
&\alpha_{1} \rho_{1}+\alpha_{2} \rho_{2}=\rho,\\
&\alpha_{1} \rho_{1} e_{1}+\alpha_{2} \rho_{2} e_{2}=\rho e,\\
\end{aligned}
\end{equation}
Derived in \cite{Shyue01}, under the isobaric assumption the mixture Gr\"{u}neisen coefficient and $p_{\infty,k}$, 
 $e_{\infty,k}$ can be expressed as
\begin{equation}
\label{eq:mixturesEOS} 
\begin{aligned}
&\frac{\alpha_{1}}{\Gamma_{1}(\rho_{1})}+\frac{\alpha_{2}}{\Gamma_{2}(\rho_{2})}=\frac{1}{\Gamma},\\
&\alpha_{1} \rho_{1} e_{\infty,1}(\rho_{1})+\alpha_{2} \rho_{2} e_{\infty,2}(\rho_{2})=\rho e_{\infty},\\
&\alpha_{1}
\frac{p_{\infty,1}(\rho_{1})}{\Gamma_{1}(\rho_{1})}+\alpha_{2}
\frac{p_{\infty,2}(\rho_{2})}{\Gamma_{2}(\rho_{2})}=
\frac{p_{\infty}(\rho)}{\Gamma(\rho)}.
\end{aligned}
\end{equation}
The mixture pressure is then calculated by 
\begin{equation}
\label{eq:mie-eos-mixture}
p = \left ( \rho e - 
\sum_{k=1}^{2} \alpha_{k} \rho_{k} e_{\infty,k}(\rho_{k})+
\sum_{k=1}^{2} \alpha_{k}
\frac{p_{\infty,k}(\rho_{k})}{\Gamma_{k}(\rho_{k})}
\right ) \bigg/ \sum_{k=1}^{2} \frac{\alpha_{k}}
{\Gamma_{k}(\rho_{k})}.
\end{equation}
It should be noted that the mixing rule of Eq.(\ref{eq:mixturesEOS}) and Eq.(\ref{eq:mie-eos-mixture}) ensure that the mixed pressure is free of spurious osclillations, which is particularly important to prevent the spurious pressure oscillations across the material interfaces \cite{Abgrall96,Abgrall01,Shyue98,Shyue01,Shyue99}. 

\section{Numerical methods \label{sec:methods}}
For the sake of simplicity, we introduce the numerical method in one dimension. Our numerical method can be extended to the multidimensions on structured grids directly in dimension-wise reconstruction fashion. We will first review the finite volume method in the wave propagation form  \cite{wave1} used in this work and then give details about the new MUSCL-THINC-BVD reconstruction scheme. 

\subsection{Wave propagation method}
We rewrite the one dimensional quasi-conservative five-equation model  (\ref{eq:5eqns}) as
\begin{equation}
	\label{eq:model-nd}
	\frac{\partial \textbf{q}}{\partial t} + \frac{\partial f(\textbf{q})}{\partial x}  +  B(\textbf{q}) \frac{\partial \textbf{q}}{\partial x} = 0,
\end{equation}
where the vectors of physical variables $\textbf{q}$ and flux functions $\textbf{f}$ are
\begin{equation}
\begin{aligned}
\textbf{q} & = \left ( \alpha_{1} \rho_{1}, \alpha_{2} \rho_{2}, 
\rho u, E, \alpha_{1} \right )^{T}, \\
\textbf{f} & = \left ( \alpha_{1} \rho_{1} u, \alpha_{2} \rho_{2} u, 
\rho uu + p,Eu + pu, 0 \right )^{T}, 
\end{aligned}
\end{equation}
respectively. The matrix $B$ is defined as
\begin{equation}
\begin{aligned}
B & = \mbox{diag}
\left ( 0, 0, 0, 0, u \right ),
\end{aligned}
\end{equation}
where $u$ denotes the velocity component in $x$ direction. 

We divide the computational domain into $N$ non-overlapping cell elements, ${\mathcal C}_{i}: x \in [x_{i-1/2},x_{i+1/2} ]$, $i=1,2,\ldots,N$, with a uniform grid with the spacing $\Delta x=x_{i+1/2}-x_{i-1/2}$. For a standard finite volume method, the volume-integrated average value $\bar{\textbf{q}}_{i}(t)$ in the cell $C_{i}$ is defined as
\begin{equation}
   \bar{\textbf{q}}_{i}(t) 
\approx \frac{1}{\Delta x} \int_{x_{i-1/2}}^{x_{i+1/2}}
\textbf{q}(x,t) \; dx.
\end{equation}  
Denoting all the spatial discretization terms in \eqref{eq:model-nd} by  $\mathcal{L}(\bar{\textbf{q}}(t))$,  
 the semi-discrete version of the finite volume formulation  can be expressed as a system of ordinary differential equations (ODEs)
\begin{equation}
 \label{eq:semi-discrete-1d-eq}
\frac{\partial \bar{\textbf{q}}(t)}{\partial t}  = 
\mathcal{L} \left (\bar{\textbf{q}}(t) \right ).
\end{equation}
In the wave-propagation method,  the spatial discretization for cell $C_{i}$ is computed by 
\begin{equation}
 \label{eq:spatial-discrete-1d}
\mathcal{L}\left( \bar{\textbf{q}}_{i}(t) \right ) =
-\frac{1}{\Delta x} \left ( 
\mathcal{A}^{+} \Delta \textbf{q}_{i-1/2} + 
\mathcal{A}^{-} \Delta \textbf{q}_{i+1/2} +
\mathcal{A} \Delta \textbf{q}_{i} \right )
\end{equation}	
where  $\mathcal{A}^{+} \Delta \textbf{q}_{i-1/2}$ and
$\mathcal{A}^{-} \Delta \textbf{q}_{i+1/2}$,
are the right- and left-moving fluctuations,
respectively, which enter into the grid cell, and
$\mathcal{A} \Delta \textbf{q}_{i}$ is
the total fluctuation within $C_{i}$. We need to solve Riemann problems to determine these fluctuations. The right- and left-moving fluctuations can be calculated by
\begin{equation}
 \mathcal{A}^{\pm} \Delta \textbf{q}_{i-1/2} =
\sum_{k = 1}^{3}
\left [s^{k} 
\left( \textbf{q}_{i-1/2}^{L},\textbf{q}_{i-1/2}^{R} \right )
\right]^{\pm}
\mathcal{W}^{k} \left( \textbf{q}_{i-1/2}^{L},\textbf{q}_{i-1/2}^{R} \right ),
\end{equation} 
where moving speeds $s^{k}$ and the jumps $\mathcal{W}^{k}$ ($k=1,2,3$) of three propagating discontinuities can be solved by Riemann solver \cite{Riemann} with the reconstructed values $\textbf{q}_{i-1/2}^{L}$ and $\textbf{q}_{i-1/2}^{R}$ computed from the reconstruction functions $\tilde{\textbf{q}}_{i-1}(x)$ and $\tilde{\textbf{q}}_{i}(x)$ to the left and right sides of cell edge $x_{i-1/2}$, respectively. Similarly, the total fluctuation can be determined by 
\begin{equation}
 \label{eq:tfluct-1d}
\mathcal{A} \Delta \textbf{q}_{i} =
\sum_{k = 1}^{3}
\left [s^{k} 
\left( \textbf{q}_{i-1/2}^{R},\textbf{q}_{i+1/2}^{L} \right )
\right]^{\pm}
\mathcal{W}^{k} \left( \textbf{q}_{i-1/2}^{R},\textbf{q}_{i+1/2}^{L} \right )
\end{equation}
We will describe with details about the reconstructions to get these values, $\textbf{q}_{i-1/2}^{L}$ and $\textbf{q}_{i-1/2}^{R}$, at cell boundaries in next subsection as the core part of this paper. 

In practice, given the reconstructed values $\textbf{q}_{i-1/2}^{L}$ and $\textbf{q}_{i-1/2}^{R}$,  the minimum and maximum moving speeds $s^{1}( \textbf{q}_{i-1/2}^{L},\textbf{q}_{i-1/2}^{R})$ and $s^{3}( \textbf{q}_{i-1/2}^{L},\textbf{q}_{i-1/2}^{R})$ can be estimated by HLLC Riemann solver \cite{Riemann}  as
\begin{equation}
\begin{aligned}
s^{1}=\mathrm{min}\{u_{i-1/2}^{L}-c_{i-1/2}^{L},u_{i-1/2}^{R}-c_{i-1/2}^{R}\},\\
s^{3}=\mathrm{max}\{u_{i-1/2}^{L}+c_{i-1/2}^{L},u_{i-1/2}^{R}+c_{i-1/2}^{R}\},
\end{aligned}
\end{equation}
where $c_{i-1/2}^{L}$ and $c_{i-1/2}^{R}$ are sound speeds calculated by reconstructed values  
$\textbf{q}_{i-1/2}^{L}$ and $\textbf{q}_{i-1/2}^{R}$ respectively. Then  the  speed of the middle wave is estimated by 
\begin{equation}
s^{2}=\dfrac{p_{i-1/2}^{R}-p_{i-1/2}^{L}+\rho_{i-1/2}^{L} u_{i-1/2}^{L} (s^{1}-u_{i-1/2}^{L})-\rho_{i-1/2}^{R} u_{i-1/2}^{R} (s^{3}-u_{i-1/2}^{R})}{\rho_{i-1/2}^{L} (s^{1}-u_{i-1/2}^{L})-\rho_{i-1/2}^{R} (s^{3}-u_{i-1/2}^{R})}.
\end{equation}

 The left-side intermediate state variables $\textbf{q}_{i-1/2}^{*L}$ is evaluated by
\begin{equation}
\textbf{q}_{i-1/2}^{*L}=\dfrac{(u_{i-1/2}^{L}-s^{1}) \textbf{q}_{i-1/2}^{L}+(p_{i-1/2}^{L}\textbf{n}_{i-1/2}^{L}-p_{i-1/2}^{*}\textbf{n}_{i-1/2}^{*} )}{s^{2}-s^{1}}
\end{equation}
where the vector $\textbf{n}_{i-1/2}^{L}=(0,0,1,u_{i-1/2}^{L},0)$, $\textbf{n}_{i-1/2}^{*}=(0,0,1,s^{2},0)$ and the intermediate pressure may be estimated as
\begin{equation}
p_{i-1/2}^{*}=\rho_{i-1/2}^{L} (u_{i-1/2}^{L}-s^{1})(u_{i-1/2}^{L}-s^{2})+p_{i-1/2}^{L}=\rho_{i-1/2}^{R} (u_{i-1/2}^{R}-s^{1})(u_{i-1/2}^{R}-s^{2})+p_{i-1/2}^{R}.
\end{equation}

Analogously, the right-side intermediate state variables $\textbf{q}_{i-1/2}^{*R}$ is 
\begin{equation}
\textbf{q}_{i-1/2}^{*R}=\dfrac{(u_{i-1/2}^{R}-s^{3}) \textbf{q}_{i-1/2}^{R}+(p_{i-1/2}^{R}\textbf{n}_{i-1/2}^{R}-p_{i-1/2}^{*}\textbf{n}_{i-1/2}^{*} )}{s^{2}-s^{3}}
\end{equation}

Then we calculate the jumps $\mathcal{W}^{k}( \textbf{q}_{i-1/2}^{R},\textbf{q}_{i+1/2}^{L})$ as
\begin{equation}
\begin{aligned}
\mathcal{W}^{1}=\textbf{q}_{i-1/2}^{*L}-\textbf{q}_{i-1/2}^{L},\\
\mathcal{W}^{2}=\textbf{q}_{i-1/2}^{*R}-\textbf{q}_{i-1/2}^{*L},\\
\mathcal{W}^{3}=\textbf{q}_{i-1/2}^{R}-\textbf{q}_{i-1/2}^{*R}.
\end{aligned}
\end{equation}

Given the spatial discretization, we employ three-stage third-order SSP (Strong Stability-Preserving) Runge-Kutta scheme \cite{ssp}
\begin{equation}
 \begin{aligned}
 &  \bar{\textbf{q}}^{\ast}  = \bar{\textbf{q}}^{n} + 
\Delta t \mathcal{L} \left( \bar{\textbf{q}}^{n} \right ), \\
&\bar{\textbf{q}}^{\ast \ast}  = \frac{3}{4} \bar{\textbf{q}}^{n} + 
\frac{1}{4} \bar{\textbf{q}}^{\ast} +
\frac{1}{4} \Delta t \mathcal{L} \left ( \bar{\textbf{q}}^{\ast} \right ), \\
&\bar{\textbf{q}}^{n+1}  = \frac{1}{3} \bar{\textbf{q}}^{n} + \frac{2}{3} \bar{\textbf{q}}^{\ast} +
\frac{2}{3} 
\Delta t \mathcal{L}\left ( \bar{\textbf{q}}^{\ast\ast} \right ),
 \end{aligned}
\end{equation}
 to solve the time evolution ODEs, where $\bar{\textbf{q}}^{\ast}$ and $\bar{\textbf{q}}^{\ast \ast}$ denote the intermediate values at the sub-steps.

\subsection{MUSCL-THINC-BVD reconstruction}
In the previous subsection, we left the boundary values, $\textbf{q}_{i-1/2}^{L}$ and $\textbf{q}_{i-1/2}^{R}$, to be determined, which are presented in this subsection.  We denote any single variable for reconstruction by $q$ , which can be primitive variable, conservative variable or characteristic variable.   

The values $q_{i-1/2}^{L}$ and $q_{i+1/2}^{R}$ at cell boundaries are computed from the piecewise reconstruction functions $\tilde{q}_{i}(x)$ in cell $C_i$. In the present work, we designed the  MUSCL-THINC-BVD reconstruction scheme to capture both smooth and nonsmooth solutions. The BVD algorithm makes use of the MUSCL scheme \cite{Van_Leer} and the THINC scheme \cite{xiao_thinc2} as the candidates for spatial reconstruction. 

In the MUSCL scheme,  a piecewise linear function is constructed from the volume-integrated average values $\bar{q}_{i}$, which reads
\begin{equation}
\tilde{q}_{i}(x)^{MUSCL}=\bar{q}_{i}+\sigma_{i}(x-x_{i})
\end{equation} 
where $x \in [x_{i-1/2},x_{i+1/2}]$ and $\sigma_{i}$ is the slope defined at the cell center $x_{i}=\frac{1}{2}(x_{i-1/2}+x_{i+1/2})$. To prevent numerical oscillation, a slope limiter \cite{Van_Leer,finite} is used to get numerical solutions  satisfying the TVD property. We denote the reconstructed value at cell boundaries from MUSCL reconstruction as $q_{i-1/2}^{R,MUSCL}$ and $q_{i+1/2}^{L,MUSCL}$. 
The MUSCL scheme, in spite of popular use in various numerical models, has excessive numerical dissipation and tends to smear out flow structures, which might be a fatal drawback in simulating interfacial multiphase flows. 

Being another reconstruction candidate, the THINC \cite{xiao_thinc,xiao_thinc2} uses the hyperbolic tangent function, which is a differentiable and monotone function that fits well a step-like discontinuity. The THINC reconstruction function is written as
\begin{equation}
\tilde{q}_{i}(x)^{THINC}=\bar{q}_{min}+\dfrac{\bar{q}_{max}}{2} \left(1+\theta~\tanh \left(\beta \left(\dfrac{x-x_{i-1/2}}{x_{i+1/2}-x_{i-1/2}}-\tilde{x}_{i}\right)\right)\right),
\end{equation} 
where $\bar{q}_{min}=min(\bar{q}_{i-1},\bar{q}_{i+1})$, $\bar{q}_{max}=max(\bar{q}_{i-1},\bar{q}_{i+1})-\bar{q}_{min}$ and $\theta=sgn(\bar{q}_{i+1}-\bar{q}_{i-1})$. The jump thickness is controlled by parameter $\beta$. In our numerical tests shown later a constant value of $\beta=1.6$ is used. The unknown $\tilde{x}_{i}$, which represents the location of the jump center, is computed from $ \bar{q}_{i} = \frac{1}{\Delta x} \int_{x_{i-1/2}}^{x_{i+1/2}} \tilde{q}_{i}(x)^{THINC} \; dx$. Then the reconstructed values at cell boundaries by THINC function can be expressed by 
\begin{equation}
\label{thniclr}
 \begin{aligned}
&q_{i+1/2}^{L,THINC}=\bar{q}_{min}+\dfrac{\bar{q}_{max}}{2} \left(1+\theta \dfrac{\tanh(\beta)+A}{1+A~\tanh(\beta)}\right)\\
&q_{i-1/2}^{R,THINC}=\bar{q}_{min}+\dfrac{\bar{q}_{max}}{2} \left(1+\theta~ A\right)
 \end{aligned}
\end{equation}
where $A=\frac{B/\cosh(\beta)-1}{\tanh(\beta)}$ and $B=\exp(\theta~\beta(2~C-1))$, where $C=\dfrac{\bar{q}_{i}-\bar{q}_{min}+\epsilon}{\bar{q}_{max}+\epsilon}$ and $\epsilon=10^{-20}$ is a mapping factor to project the physical fields onto $[0,1]$.

The final effective reconstruction function is determined by the BVD algorithm \cite{Sun,Xie}, which choose the reconstruction function between $\tilde{q}_{i}(x)^{MUSCL}$ and $\tilde{q}_{i}(x)^{THINC}$ so that the variations of the reconstructed values at cell boundaries are minimized. BVD algorithm prefers the THINC reconstruction $\tilde{q}_{i}(x)^{THINC}$ within a cell where a discontinuity exists. It is sensible that the THINC reconstruction should only be employed when a discontinuity is detected. In practice, a cell where a discontinuity may exist can be identified by the following conditions
\begin{equation}
\label{eq:thincC}
 \begin{aligned}
 &\delta<C<1-\delta,\\
 &(\bar{q}_{i+1}-\bar{q}_{i})(\bar{q}_{i}-\bar{q}_{i-1})>0,
 \end{aligned}
\end{equation}
where $\delta$ is a small positive (e.g.,$10^{-4}$). 
 
 In summary, the effective reconstruction function of MUSCL-THINC-BVD scheme reads 
 \begin{equation}
 \tilde{q}_{i}(x)^{BVDl}=\left\{
 \begin{array}{l}
 \tilde{q}_{i}(x)^{THINC}~~~\mathrm{if}~\delta<C<1-\delta,~\mathrm{and}~ (\bar{q}_{i+1}-\bar{q}_{i})(\bar{q}_{i}-\bar{q}_{i-1})>0,~\mathrm{and}~TBV_{i,min}^{THINC}<TBV_{i,min}^{MUSCL}\\
 \tilde{q}_{i}(x)^{MUSCL}~~~\mathrm{otherwise}
 \end{array}
 \right..
 \end{equation}
 where the minimum value of total boundary variation (TBV) $TBV_{i,min}^{P}$, for reconstruction function $P =THINC~or~MUSCL$, is defined as
 \begin{equation}
 \begin{aligned}
 TBV_{i,min}^{P}=min(&|q_{i-1/2}^{L,MUSCL}-q_{i-1/2}^{R,P}|+|q_{i+1/2}^{L,P}-q_{i+1/2}^{R,MUSCL}|,|q_{i-1/2}^{L,THINC}-q_{i-1/2}^{R,P}|+|q_{i+1/2}^{L,P}-q_{i+1/2}^{R,THINC}|,
 \\&|q_{i-1/2}^{L,MUSCL}-q_{i-1/2}^{R,P}|+|q_{i+1/2}^{L,P}-q_{i+1/2}^{R,THINC}| ,|q_{i-1/2}^{L,THINC}-q_{i-1/2}^{R,P}|+|q_{i+1/2}^{L,P}-q_{i+1/2}^{R,MUSCL}|) .
 \end{aligned}
 \end{equation}
  
\begin{figure}[h]
	\centering{}
	\includegraphics[scale=0.6]{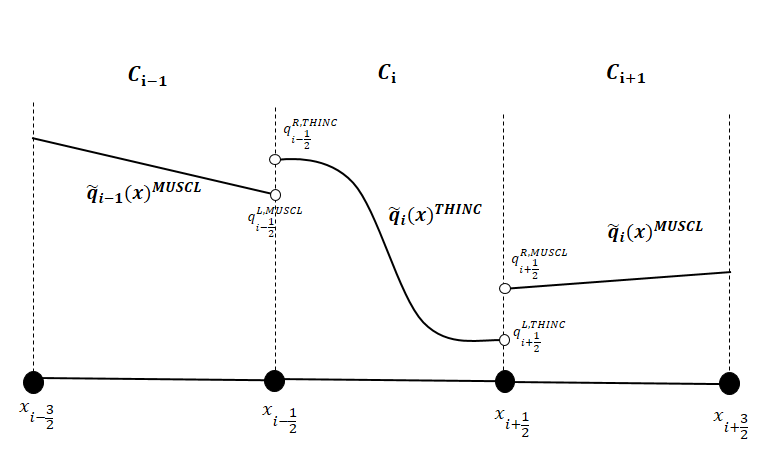}
	\protect\caption{Illustration of one possible situation corresponding to $|q_{i-1/2}^{L,MUSCL}-q_{i-1/2}^{R,THINC}|+|q_{i+1/2}^{L,THINC}-q_{i+1/2}^{R,MUSCL}|$ when calculating $TBV_{i,min}^{THINC}$.\label{fig:BV} }
\end{figure} 
Thus, THINC reconstruction function will be employed in the targeted cell if the minimum TBV value of THINC is smaller than that of MUSCL. In Fig.~\ref{fig:BV}, we illustrate one possible situation corresponding to $|q_{i-1/2}^{L,MUSCL}-q_{i-1/2}^{R,THINC}|+|q_{i+1/2}^{L,THINC}-q_{i+1/2}^{R,MUSCL}|$ when evaluating $TBV_{i,min}^{THINC}$. As stated in \cite{Sun}, the BVD algorithm will realize the polynomial interpolation for smooth solution while for discontinuous solution a step like function will be preferred. 

As shown in numerical tests in this paper, discontinuities including material interface can be resolved by the MUSCL-THINC-BVD scheme with substantially reduced numerical dissipation in comparison with other existing methods. The material interface can be captured sharply and any extra step, like anti-diffusion or other artificial interface sharpening techniques used in the existing works \cite{Gao,Shyue2014,Shukla1,Shukla2}, is not needed here.  More importantly, the  MUSCL-THINC-BVD scheme is applied not only the volume fraction but also to other all physical variables, which automatically leads to the consistent reconstructions among the physical fields. As observed in our numerical results, no suprious numerical oscillation is generated in vicinity of material interfaces.  It is usualy not trival to other anti-diffusion or artificial compression methods aforementioned.  For example, in \cite{Tiwari,Shukla2,Shyue2014} anti-diffusion post-processing steps are required to adjust other state variables across interfaces to get around the oscillations. 

As discussed in \cite{Johnsen1,Johnsen2,weno1,weno2}, when high-order reconstructions, such as MUSCL or WENO shemes, are applied, special attention must be paid to decide which physical variables should be reconstructed. It is concluded that one should implement high-order reconstructions to primitive variables or characteristic variables to prevent numerical oscillations in velocity and pressure across material interfaces. However, reconstructing conservative variables or flux functions may cause numerical oscillations. We show in the rest part of this section that THINC reconstruction ensures the consistency among the reconstructed variables across material interface even if the reconstruction is conducted for the conservative variables.

We consider one-dimensional interface only problem where initial condition consists of constant velocity $u=u_{0}$, uniform pressure $p=p_{0}$ and constant phasic densities $\rho_{1}=\rho_{10}$ and $\rho_{2}=\rho_{20}$. Across a material interface, other variables,  such as mixture densities $\rho$, mass fraction $\alpha_{1} \rho_{1}$ and volume fraction $\alpha_{1}$ have jumps. Without loss of generality, a positive velocity $u=u_{0}>0$ is considered here. Then the fluctuations for cell $C_{i}$ can be calculated as
\begin{equation}
\label{eq:interface1}
\begin{aligned}
&\mathcal{A}^{-} \Delta \textbf{q}_{i+1/2}  = 0, \\
&\mathcal{A}^{+} \Delta \textbf{q}_{i-1/2}  = 
u_{0} \left ( \textbf{q}_{i-1/2}^{R} - \textbf{q}_{i-1/2}^{L} \right ), \\
&\mathcal{A} \Delta \textbf{q}_{i}  = 
u_{0} \left ( \textbf{q}_{i+1/2}^{L} - \textbf{q}_{i-1/2}^{R} \right ).
\end{aligned}
\end{equation}
By using Eq. (\ref{eq:spatial-discrete-1d}) and Euler one-step forward time scheme, the cell average $\bar{\textbf{q}}_{i}^{n}$ can be updated by 
\begin{equation}
\bar{\textbf{q}}_{i}^{n+1} = \bar{\textbf{q}}_{i}^{n} - 
\frac{\Delta t}{\Delta x}
u_{0} \left ( \textbf{q}_{i+1/2}^{L} - \textbf{q}_{i-1/2}^{L} \right ), 
\end{equation}
or a component form, 
\begin{equation}
\label{eq:meth-intf0}
\left [
\begin{matrix}
\alpha_{1} \rho_{1} \\
\alpha_{2} \rho_{2} \\
\rho u \\
E \\
\alpha_{1}
\end{matrix}
\right ]_{i}^{n+1} =
\left [
\begin{matrix}
\alpha_{1} \rho_{1} \\
\alpha_{2} \rho_{2} \\
\rho u \\
E \\
\alpha_{1} 
\end{matrix}
\right ]_{i}^{n} -
\frac{\Delta t}{\Delta x}
u_{0}
\left [
\begin{matrix}
\left ( \alpha_{1} \rho_{1} \right )_{i+1/2}^{L}-
\left ( \alpha_{1} \rho_{1} \right )_{i-1/2}^{L}
\\
\left ( \alpha_{2} \rho_{2} \right )_{i+1/2}^{L}-
\left ( \alpha_{2} \rho_{2} \right )_{i-1/2}^{L}
\\
u_{0} \left ( \rho_{i+1/2}^{L}-\rho_{i-1/2}^{L}
\right )
\\
E_{i+1/2}^{L}-E_{i-1/2}^{L}
\\ 
\left (\alpha_{1} \right )_{i+1/2}^{L}-
\left (\alpha_{1} \right )_{i-1/2}^{L}
\end{matrix}
\right ].
\end{equation}
We denote the reconstruction operator to compute $q_{i+1/2}^{L}$ by  $\mathcal{D}_{i+\frac{1}{2}}(q)^{L}$. 

As shown below, when we implement the reconstructions to the conservative variables, spurious oscillations may be generated in velocity $u$. To facilitate discussions, we address the  consistency in velocity $u$ by   
 
\textbf{Definition 1.} \textit{The reconstruction is $u$-consistent if the numerical results from \eqref{eq:meth-intf0} satisfy   $u^{n+1}=u^{n}$ for isolated material interface where velocity $u$ and pressure $p$ are uniform.}

In the present model, in order to calculate velocity $u_{i}^{n+1}$, we first compute density from the two conservation equations for phasic densities, 
\begin{equation}
\rho_{i}^{n+1}=\rho_{i}^{n}-\frac{\Delta t}{\Delta x}u_{0}\left((\mathcal{D}_{i+\frac{1}{2}}(\alpha_{1} \rho_{1})^{L}+\mathcal{D}_{i+\frac{1}{2}}(\alpha_{2} \rho_{2})^{L})-(\mathcal{D}_{i-\frac{1}{2}}(\alpha_{1} \rho_{1})^{L}+\mathcal{D}_{i-\frac{1}{2}}(\alpha_{2} \rho_{2})^{L})\right).
\end{equation}
From the momentum equation, we have 
\begin{equation}
(\rho u)_{i}^{n+1}=(\rho u)_{i}^{n}-\frac{\Delta t}{\Delta x}u_{0}\left(\mathcal{D}_{i+\frac{1}{2}}(\rho u)^{L}-\mathcal{D}_{i-\frac{1}{2}}(\rho u)^{L}\right)=\rho_{i}^{n} u_{0}-\frac{\Delta t}{\Delta x}u_{0}u_{0}\left(\mathcal{D}_{i+\frac{1}{2}}(\rho)^{L}-\mathcal{D}_{i-\frac{1}{2}}(\rho)^{L}\right),
\end{equation}
where we assume the operator should satisfy $\mathcal{D}_{j}(mX)=m\mathcal{D}_{j}(X)$ in which $X$ represents the reconstructed variable, $m$ is constant, then the velocity $u^{n+1}$ is retrieved by 
\begin{equation}
\label{uequ}
u_{i}^{n+1}=u_{0}\dfrac{\rho_{i}^{n}-\frac{\Delta t}{\Delta x}u_{0}\left(\mathcal{D}_{i+\frac{1}{2}}(\rho)^{L}-\mathcal{D}_{i-\frac{1}{2}}(\rho)^{L}\right)}{\rho_{i}^{n}-\frac{\Delta t}{\Delta x}u_{0}\left((\mathcal{D}_{i+\frac{1}{2}}(\alpha_{1} \rho_{1})^{L}+\mathcal{D}_{i+\frac{1}{2}}(\alpha_{2} \rho_{2})^{L})-(\mathcal{D}_{i-\frac{1}{2}}(\alpha_{1} \rho_{1})^{L}+\mathcal{D}_{i-\frac{1}{2}}(\alpha_{2} \rho_{2})^{L})\right)}.
\end{equation}
It is clear that  the $u$-consistent condition requires 
\begin{equation} 
\label{u-const}
\mathcal{D}_{i\pm\frac{1}{2}}(\rho)^{L}=(\mathcal{D}_{i\pm\frac{1}{2}}(\alpha_{1} \rho_{1})^{L}+\mathcal{D}_{i\pm\frac{1}{2}}(\alpha_{2} \rho_{2})^{L})
\end{equation}
 to maintain $u_{i}^{n+1}=u_{0}$.  I.e. the reconstruction operator $\mathcal{D}_{i\pm\frac{1}{2}}(\rho )^{L}$ should be consistent with $\mathcal{D}_{i\pm\frac{1}{2}}(\alpha_{1} \rho_{1})^{L}$ and $\mathcal{D}_{i\pm\frac{1}{2}}(\alpha_{2} \rho_{2})^{L}$ at cell faces $i\pm\frac{1}{2}$ to ensure $\rho_{i\pm\frac{1}{2}}^{L}=(\alpha_{1} \rho_{1})_{i\pm\frac{1}{2}}^{L}+(\alpha_{2} \rho_{2})_{i\pm\frac{1}{2}}^{L}$. 

Concerning $u$-consistent property, we have 

\textbf{Proposition 1.} \textit{All schemes satisfy $u$-consistent condition if reconstruction are conducted in terms of primitive variables.}

\textbf{Proof.} It is straightforward that $u^{n+1}=u^{n}$ since the reconstruction is conducted with $u$, which has also been discussed in \cite{Abgrall96,weno2}. $\square$

\textbf{Proposition 2.} \textit{Piecewise constant reconstruction scheme for conservative variables satisfy $u$-consistent condition}.

\textbf{Proof.} 
 
 For piecewise constant reconstruction, it is obvious that above condition \eqref{u-const} can be satisfied, as $\rho_{i-\frac{1}{2}}^{L}=\rho_{i-1}=(\alpha_{1} \rho_{1})_{i-1}+(\alpha_{2} \rho_{2})_{i-1}=(\alpha_{1} \rho_{1})_{i-\frac{1}{2}}^{L}+(\alpha_{2} \rho_{2})_{i-\frac{1}{2}}^{L}$ and $\rho_{i+\frac{1}{2}}^{L}=\rho_{i}=(\alpha_{1} \rho_{1})_{i}+(\alpha_{2} \rho_{2})_{i}=(\alpha_{1} \rho_{1})_{i+\frac{1}{2}}^{L}+(\alpha_{2} \rho_{2})_{i+\frac{1}{2}}^{L}$.$\square$
 
\textbf{Proposition 3.} \textit{All linear reconstruction schemes for conservative variables satisfy $u$-consistent condition}.

\textbf{Proof.} 

For general linear schemes, the reconstructed values at cell boundaries can be expressed as linear combinations of the cell average values, $\bar{q}_{l}$, on the stencils, ${\mathcal C}_l, \ l=i-l',\cdots, i+l''$. That is  
\begin{equation}
\mathcal{D}_{i+\frac{1}{2}}(q)^{L}=\sum_{l}\chi_{l}\bar{q}_{l},
\end{equation}
where the coefficients $\chi_{l}$ are the same for $\rho$, $\alpha_{1} \rho_{1}$ and   $\alpha_{2} \rho_{2}$. With the conclusion in proposition 2  given above for piecewise constant reconstruction, we know that $\mathcal{D}_{i+\frac{1}{2}}(\rho)^{L}=(\mathcal{D}_{i+\frac{1}{2}}(\alpha_{1} \rho_{1})^{L}+\mathcal{D}_{i+\frac{1}{2}}(\alpha_{2} \rho_{2})^{L})$, which leads to $u_{i}^{n+1}=u_{0}$ from Eq.~(\ref{uequ}). $\square$

\textbf{Proposition 4.} \textit{Non-linear schemes for conservative variables may not satisfy $u$-consistent condition}.

\textbf{Proof.} Taking 5th order WENO scheme as an example, the reconstructed values at cell faces are a combination of three third order linear schemes through nonlinear weights. The reconstructed value can be expressed as
\begin{equation}
q_{i+1/2}^{L}=\sum_{k=1}^{3}w_{i+\frac{1}{2}}^{(k)}(q)^{L}\mathcal{D}_{i+\frac{1}{2}}^{(k)}(q)^{L},
\end{equation}
where $\mathcal{D}_{i+\frac{1}{2}}^{(1)}(q)^{L}$ is the third order approximation on stencil $\{ {\mathcal C}_{i-2}, {\mathcal C}_{i-1}, {\mathcal C}_{i} \}$, $\mathcal{D}_{i+\frac{1}{2}}^{(2)}(q)^{L}$ on $\{{\mathcal C}_{i-1}, {\mathcal C}_{i}, {\mathcal C}_{i+1}\}$ and $\mathcal{D}_{i+\frac{1}{2}}^{(3)}(q)^{L}$ on $\{C_{i}, C_{i+1}, C_{i+2}\}$. $w_{i+\frac{1}{2}}^{(k)}(q)^{L}$ is the nonlinear weight corresponding to the $\mathcal{D}_{i+\frac{1}{2}}^{(k)}(q)^{L}$. To satisfy $u$-consistent condition, it is necessary that 
\begin{equation}
\label{wequ}
w_{i+\frac{1}{2}}^{(k)}(\alpha_{1} \rho_{1})^{L}=w_{i+\frac{1}{2}}^{(k)}(\alpha_{2} \rho_{2})^{L}=w_{i+\frac{1}{2}}^{(k)}(\rho u)^{L}.
\end{equation}
Because the nonlinear weights are separately determined according the smoothness of the reconstructed variables, there is no guarantee that   Eq.~(\ref{wequ}) always holds.  Thus,  WENO scheme is not $u$-consistent when applied to conservative variables, which has also been reported in \cite{Johnsen1,Johnsen2}. This observation applies to all high-resolution schemes using nonlinear weights to suppress numerical oscillations. $\square$

\textbf{Proposition 5.} \textit{THINC reconstruction satisfies $u$-consistent condition.}  

\textbf{Proof.} We consider a material interface in cell $C_{i}$ with volume fraction $(\alpha_{1})_{i}=\lambda(\alpha_{1})_{i-1}+(1-\lambda)(\alpha_{1})_{i+1}$, $0<\lambda<1$, which divides the cell into two regions sharing uniform physical properties with the neighboring cells respectively.  Then, we have $(\rho u)_{i}=\lambda(\rho u)_{i-1}+(1-\lambda)(\rho u)_{i+1}$.  Without loss of generality, we assume $(\alpha_{1})_{i-1}>(\alpha_{1})_{i+1}$ and $\rho_{1}>\rho_{2}$, the reconstructed values of phasic densities at cell face $i+\frac{1}{2}$ are
\begin{equation}
\begin{aligned}
( \alpha_{1} \rho_{1})_{i+1/2}^{L}=(\alpha_{1} \rho_{1})_{i+1}+\dfrac{(\alpha_{1} \rho_{1})_{i-1}-(\alpha_{1} \rho_{1})_{i+1}}{2}\left(1- \dfrac{\tanh(\beta)+A_{1}}{1+A_{1}~\tanh(\beta)}\right); \\
( \alpha_{2} \rho_{2})_{i+1/2}^{L}=(\alpha_{2} \rho_{2})_{i-1}+\dfrac{(\alpha_{2} \rho_{2})_{i+1}-(\alpha_{2} \rho_{2})_{i-1}}{2}\left(1+ \dfrac{\tanh(\beta)+A_{2}}{1+A_{2}~\tanh(\beta)}\right). 
\end{aligned}
\end{equation}
Recall \eqref{thniclr}, we get  $C_{1}=\lambda$ and $C_{2}=(1-\lambda)$, which leads to $B_{1}=\exp(\beta(1-2\lambda))=B_{2}$ and  $A_{1}=A_{2}=A$. 

From the momentum equation, we have
\begin{equation}
\rho_{i+1/2}^{L}=\rho_{i+1}+\dfrac{\rho_{i-1}-\rho_{i+1}}{2}\left(1- \dfrac{\tanh(\beta)+A_{3}}{1+A_{3}~\tanh(\beta)}\right),
\end{equation}
Again we can find $A_{3}=A$. Then, we finally get
\begin{equation}
\rho_{i+1/2}^{L}=(\alpha_{1} \rho_{1})_{i+1/2}^{L}+(\alpha_{2} \rho_{2})_{i+1/2}^{L}.
\end{equation}
Thus, $u$-consistent condition is satisfied. $\square$

We remark that the conclusion of proposition 4 applies to any reconstruction function if it is used in exactly the same form to different reconstructed fields. Being an extreme case, when $\beta$ is large enough, the THINC reconstruction build up two piecewise-constant states in the cell where discontinuities exist. It might be of physical importance in applications. 

To verify that THINC scheme satisfies $u$-consistent condition even used to reconstruct the conservative variables, we present the numerical results to the isolated interface problem similar to \cite{weno2,nonomura12,kawai12}, where the initial condition is set as follows
\begin{equation}
\left(\alpha_{1}\rho_{1},\ \alpha_{2}\rho_{2},\ u_{0},\ p_{0},\ \alpha_{1},\ \gamma \right)=\left\{
\begin{array}{l}
\left(10,\ 0,\ 0.5,\ \frac{1}{1.4},\ 1,\ 1.6\right) \ \mathrm{for}\ 0.3 \leq x < 0.7 \\
\left(0,\ 0.5,\ 0.5,\ \frac{1}{1.4},\ 0,\ 1.4\right) \ \mathrm{else}
\end{array}
\right..
\end{equation}
The computational domain is $[0,1]$. We computed the same test with MUSCL\cite{Van_Leer}, WENO\cite{Jiang} and THINC scheme respectively for comparison. Reconstruction is conducted in terms of the conservative variables. Figure~2 shows the numerical results of WENO and THINC at time $t=0.1$ using a $200$-cell mesh with the Courant-Friedrichs-Lewy (CFL) condition $CFL = 0.5$. From the results, we can see WENO scheme produces numerical oscillations due to its non-linear property. We also compare the results computed from MUSCL and THINC in Figure~\ref{fig:balancem}. MUSCL scheme also produces numerical oscillations when reconstructing with conservative variables. The numerical oscillations generated from the MUSCL reconstruction are much smaller than that from the WENO scheme. 

Previous works suggested the use of primitive variables or characteristic variables for reconstruction, such as \cite{Abgrall96,five} for MUSCL and  \cite{weno1,weno2} for WENO. The THINC reconstruction provides consistent reconstruction for all types of variables, which indicates more possibility in applications. 
 
\begin{figure}
	\subfigure[$p$]{\centering\includegraphics[scale=0.35]{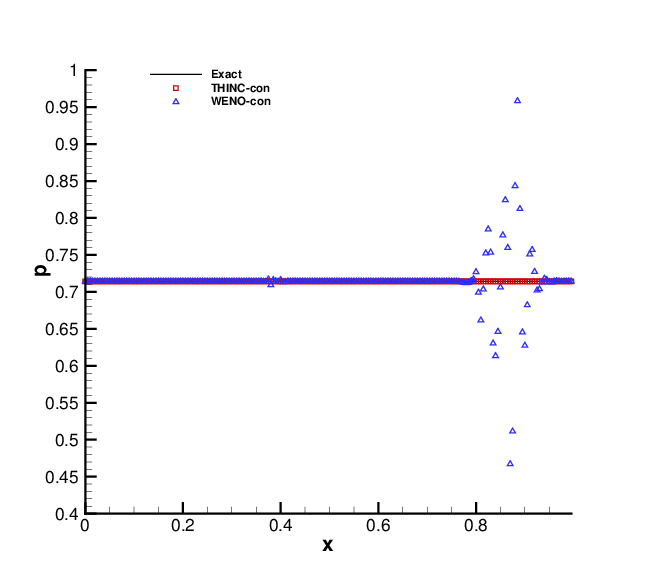}}
	\subfigure[$u$]{\centering\includegraphics[scale=0.35]{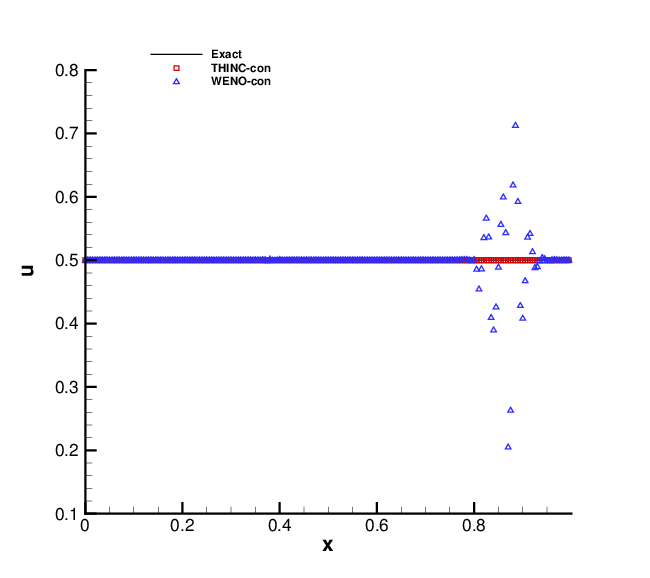}}
	\protect\caption{Comparative numerical results between WENO and THINC for conservative ones}	
\end{figure}

\begin{figure}
	\subfigure[$p$]{\centering\includegraphics[scale=0.35]{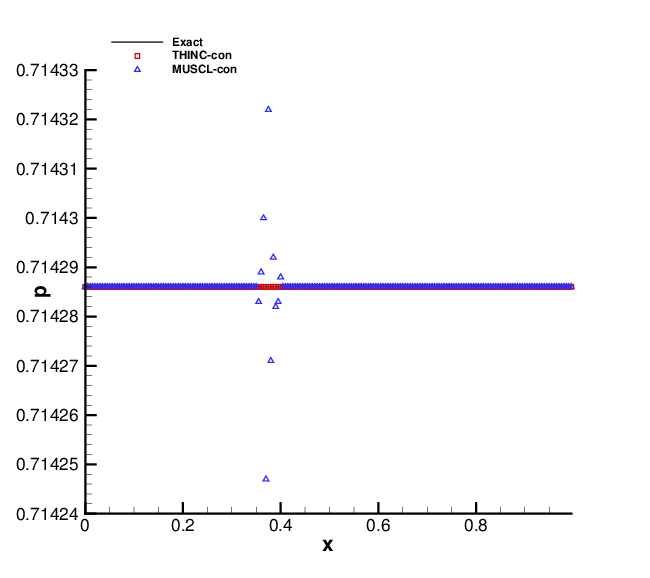}}
	\subfigure[$u$]{\centering\includegraphics[scale=0.35]{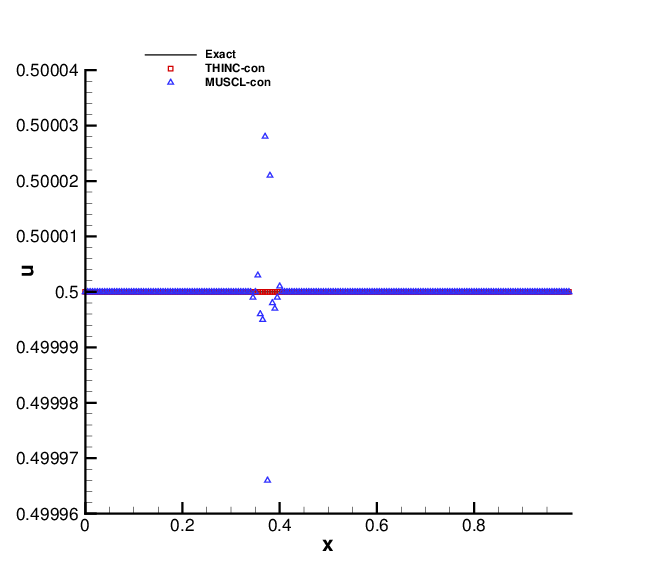}}
	\protect\caption{Same as Figure 2, but comparison between MUSCL and THINC}	
\end{figure}

\section{Numerical results \label{sec:results}}
Comparative tests in one- and two- dimensions are conducted in this section with WENO scheme and the proposed MUSCL-THINC-BVD scheme. Here we use the WENO scheme in \cite{Jiang} which is one of representative high order shock-capturing schemes. We denote it as WENO-JS in our tests. In order to reduce the numerical oscillations, the WENO reconstruction should be implemented for characteristic fields as \cite{weno1}.  

The one dimension tests were conducted with a single CPU (Intel(R) Xeon(R) CPU E5-2687W, 3.10GHZ), while two dimensional tests were conducted with a NVIDIA GTX980ti GPU. 

\subsection{Passive advection of a square liquid column \label{test1}}

To evaluate the ability of the proposed scheme to capture interface as well as to maintain the equilibrium of velocity and pressure fields, a simple interface-only problem in one dimension is considered in this test. The problem consists of a square liquid column 
in gas transported with a uniform velocity $u = u_{0} = 10^{2}$ m/s under equilibrium pressure $p= p_{0} =10^{5}$ Pa  
in a shock tube of one meter. For initial condition, liquid is set in the region of $x \in [0.4, 0.6]$m  and gas is filled elsewhere.  We set initially the volume fraction of liquid  $\alpha_{1} =1-\epsilon$ for the liquid region and $ \alpha_{1} =\epsilon$ in the gas region, and the volume fraction of gas is then   $\alpha_{2} =1- \alpha_{1} $. The small positive  $ \epsilon$ is set $10^{-8}$ in numerical tests in this paper. The densities for the liquid and gas phases are $\rho_{1}= 10^{3}$kg/m$^3$ and $\rho_{2} = 1$kg/m$^3$, respectively. 

To model the thermodynamic behavior of
liquid and gas, we use the stiffened gas equation of state where    
the material-dependent functions appeared in~(\ref{eq:mg-eos}) are
\[
\Gamma_{k} = \gamma_{k} -1, \qquad
p_{\infty,k} = \gamma_{k} \mathcal{B}_{k}, \qquad
e_{\infty,k} = 0, 
\]
with the parameter values taken in turn to be $\gamma_{1} = 4.4$,
$\mathcal{B}_{1} = 6 \times 10^{8}$Pa,
and $\gamma_{2} = 1.4$,
$\mathcal{B}_{2} = 0$
for the liquid and gas phases.

\begin{figure}
	\subfigure[$\alpha_{1}\rho_{1}$]{\centering\includegraphics[scale=0.35]{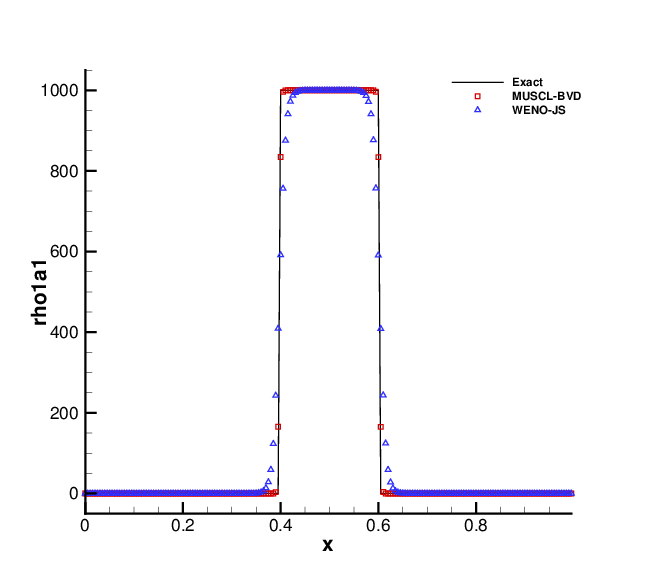}}
	\subfigure[$p$]{\centering\includegraphics[scale=0.35]{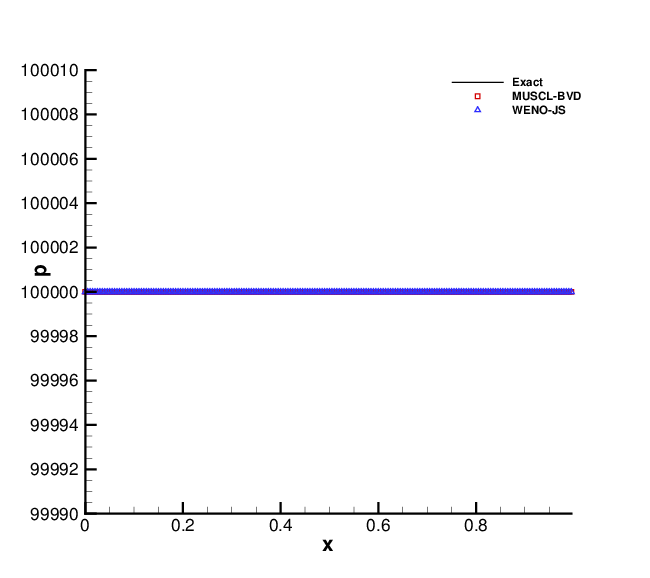}}
	\protect\caption{Numerical results for a passive advection of a square liquid column at time $t=10$ms. The solid line is the exact solution and the points shows the computed solution with $200$ mesh points obtained using different methods.  We denote the numerical result from MUSCl-THINC-BVD by MUSCL-BVD and that from \cite{Jiang} by WENO-JS. \label{fig:advection}}	
\end{figure}

The computations using WENO-JS and MUSCL-THINC-BVD are carried out respectively.
Periodic boundary condition is used on the
left and right boundaries during the computations.
Figure~\ref{fig:advection} shows numerical results of  partial density and pressure fields 
at time $t=10$ms using a $200$-cell mesh with $CFL = 0.5$.  It is obvious that MUSCl-BVD can solve the sharp interface within only two cells while WENO scheme, in spite of  high-order accuracy, excessively diffuses the interface due to the intrinsic numerical dissipation as other conventional shock capturing schemes. Meanwhile, MUSCL-THINC-BVD can retain the correct pressure equilibrium and
particle velocity without introducing spurious oscillations
across the interfaces. We have not conducted any extra procedures to sharpen the interface, which are used in other existing works to keep the steepness of the jump in volume fraction field to identify the interface. The  MUSCL-THINC-BVD reconstruction is implemented to all state variables, which remains the thermo-dynamical consistency among the physical fields. We also compare the computational cost in Table~\ref{cost}. Since it is not necessary to cast state variables to characteristic fields, the computation cost of MUSCL-THINC-BVD is about half of the WENO scheme.

\begin{table}[]
	\centering
	\caption{Comparison between WENO and MUSCL-THINC-BVD regarding to the elapse time for one dimensional tests}
	\label{cost}
	\begin{tabular}{llll}
		\hline
		& Test~\ref{test1} & Test~\ref{test2} & Test~\ref{test3} \\ \hline
		WENO-JS   & 9.90s & 3.02s & 8.90s \\
		MUSCL-THINC-BVD & 4.23s & 1.74s & 3.72s \\ \hline
	\end{tabular}
\end{table}

\subsection{Two-material impact problem \label{test2}}
Following  \cite{test2_1,test2_2}, we computed the two-phase impact benchmark problem.
At the beginning, there is a right-moving copper (phase 1) plate with
the speed $u_{1}=1500$~m/s interacting with a solid explosive (phase 2) 
at rest on the right of the plate under the uniform atmospheric condition which has pressure $p_{0} = 10^{5}$ Pa and temperature $T_{0} = 300$K throughout the domain. 
The material properties of the copper and (solid) explosive are modeled by 
the Cochran-Chan equation of state where in~(\ref{eq:mg-eos}) 
we set the same $\Gamma_{k}$ as in the stiffened gas case,
but with $p_{\infty,k}$, $e_{\infty,k}$ defined by
\begin{equation}
\label{eq:cc-eos}
\begin{aligned}
p_{\infty,k}(\rho_{k}) & = 
\mathcal{B}_{1k} \left ( \frac{\rho_{0k}}{\rho_{k}}
\right )^{-\mathcal{E}_{1k}} -
\mathcal{B}_{2k} \left ( \frac{\rho_{0k}}{\rho_{k}}
\right )^{-\mathcal{E}_{2k}}, \\
e_{\infty,k}(\rho_{k}) & = \frac{-\mathcal{B}_{1k}} 
{\rho_{0k} \left ( 1-\mathcal{E}_{1k} \right ) }
\left [ \left ( \frac{\rho_{0k}}{\rho_{k}}
\right )^{1-\mathcal{E}_{1k}}
\hskip -0.2in - 1 \right ] + \\ 
& \hskip 0.6in
\frac{\mathcal{B}_{2k}}
{\rho_{0k} \left ( 1-\mathcal{E}_{2k} \right )}
\left [
\left ( \frac{\rho_{0k}}{\rho_{k}}
\right )^{1-\mathcal{E}_{2k}} 
\hskip -0.2in - 1
\right ]- C_{vk}T_{0}. \\
\end{aligned}
\end{equation}
Here $\gamma_{k}$, $\mathcal{B}_{1k}$,
$\mathcal{B}_{2k}$, $\mathcal{E}_{1k}$, $\mathcal{E}_{2k}$,
$C_{vk}$, and $\rho_{0k}$ are material-dependent quantities,
see Table~\ref{table:test1d-2} for a typical set of
numerical values for copper and explosive considered.

\begin{table}
	\caption{Material quantities for copper ($k=1$) and
		explosive ($k=2$) in
		Cochran-Chan equation of state~(\ref{eq:cc-eos}).
	} 
	\label{table:test1d-2}
	\begin{center}
		\begin{tabular}{|c|ccccccc|} 
			\hline   
			k  & $\rho_{0k}$(kg/m$^{3}$) 
			& $\mathcal{B}_{1k}$(GPa) & $\mathcal{B}_{2k}$(GPa)
			& $\mathcal{E}_{1k}$ & $\mathcal{E}_{2k}$ 
			& $\gamma_{k}$  & $C_{vk}$     \\  \hline
			$1$ & $8900$ & $145.67$ & $147.75$ &
			$ 2.99$ & $1.99$ & $3$ & $393$J/kg$\cdot$K \\ 
			\hline
			$2$ & $1840$ & $12.87$ & $13.42$ &
			$ 4.1$ & $3.1$ & $1.93$ & 
			$1087$J/kg$\cdot$K \\ \hline
		\end{tabular}
	\end{center}
\end{table}

\begin{figure}
	\subfigure[$\alpha_{1}$]{\centering\includegraphics[scale=0.35]{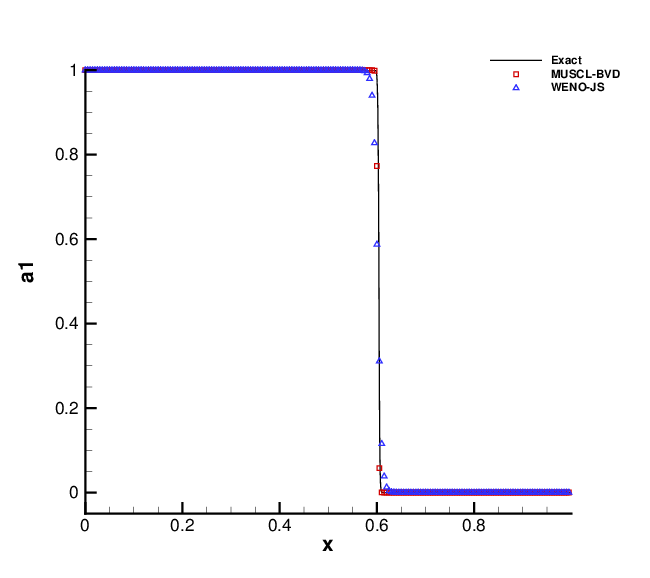}}
	\subfigure[$\alpha_{1}\rho_{1}$]{\centering\includegraphics[scale=0.35]{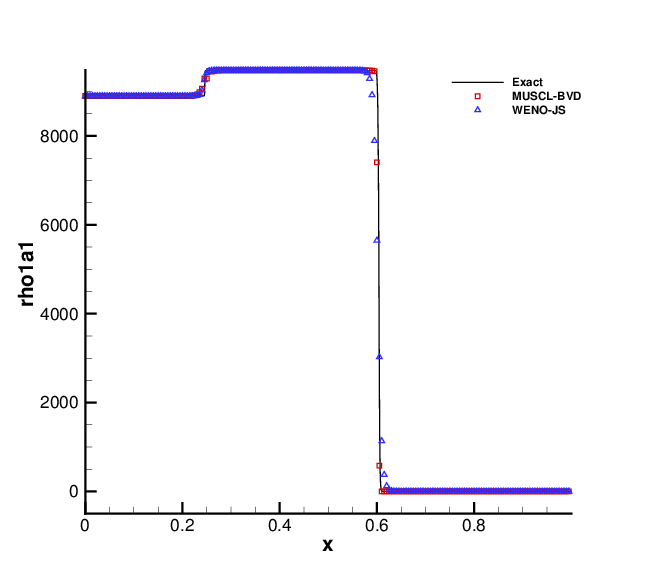}}
	\qquad{}
		\subfigure[$\alpha_{2}\rho_{2}$]{\centering\includegraphics[scale=0.35]{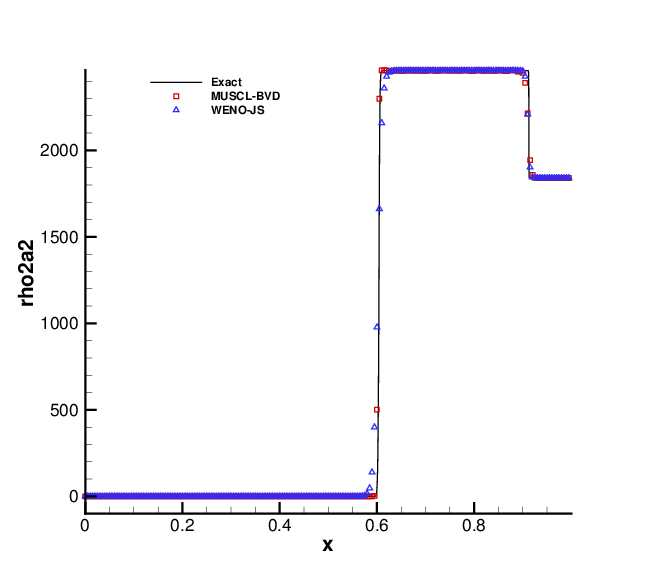}}
	\subfigure[$u$]{\centering\includegraphics[scale=0.35]{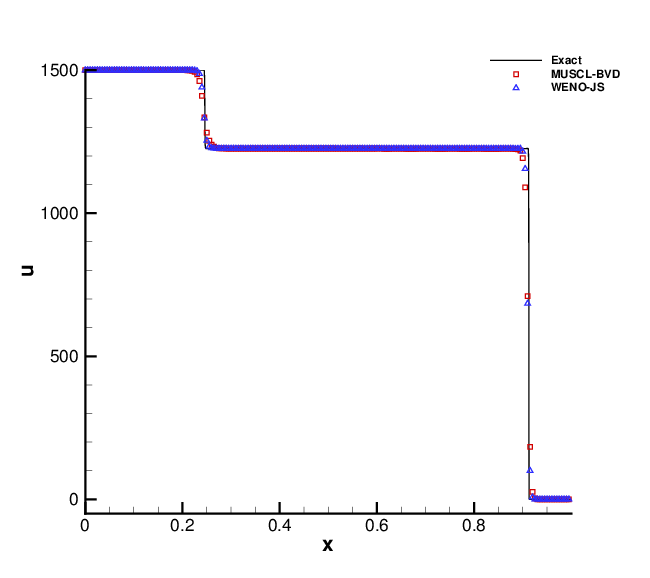}}
	\protect\caption{Numerical results for a two-phase
		(solid explosive-copper) impact problem
		at time $t= 85 \mu$s. The solid line is the fine grid solution computed on a mesh of  $5000$ cells by MUSCL, and the points
		show the solution with $200$ meshes.\label{fig:impact}}	
\end{figure}

The solution of this test is characterized by a left-moving
shock wave to the copper, a right-moving shock waves to the inert explosive,
and a material interface lying in between that separates these two different materials. We run this problem with a $200$-cell grid and $CFL=0.5$ up to  $t= 85 \mu$s. 
Figure~\ref{fig:impact} shows the results 
for the partial densities, velocity, and the copper volume fraction of both  WENO and MUSCL-THINC-BVD for comparison. Again, MUSCL-THINC-BVD can keep sharp interface without spurious numerical oscillation in velocity fields. It should be noted that due to complicated state equations, characteristic decomposition is conducted as in \cite{Gao} when implementing WENO scheme. In previous work \cite{Shyue2014}, there is a slight overshoot on the partial density 
$\alpha_{1} \rho_{1}$ on the left of the interface when using THINC method for the volume fraction. This oscillation is not observed in present study due to the global consistency in MUSCL-THINC-BVD reconstructions for all physical fields.

\subsection{Shock interface interaction problem \label{test3}}
\begin{figure}
	\subfigure[$\alpha_{1}$]{\centering\includegraphics[scale=0.35]{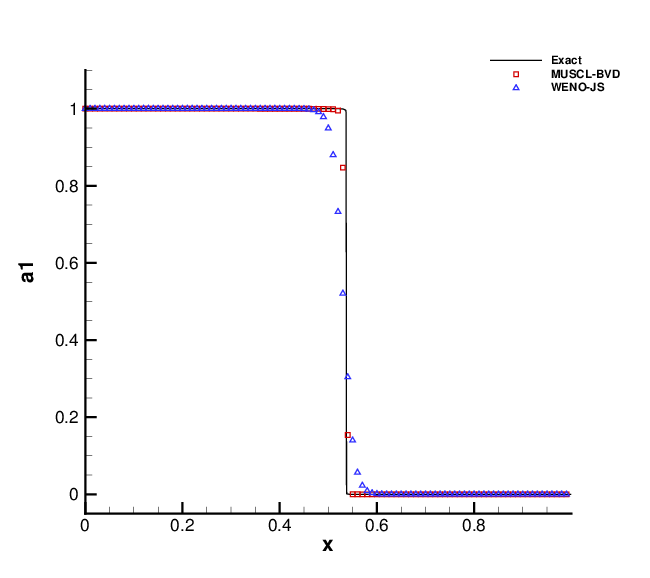}}
	\subfigure[$\alpha_{2}\rho_{2}$]{\centering\includegraphics[scale=0.35]{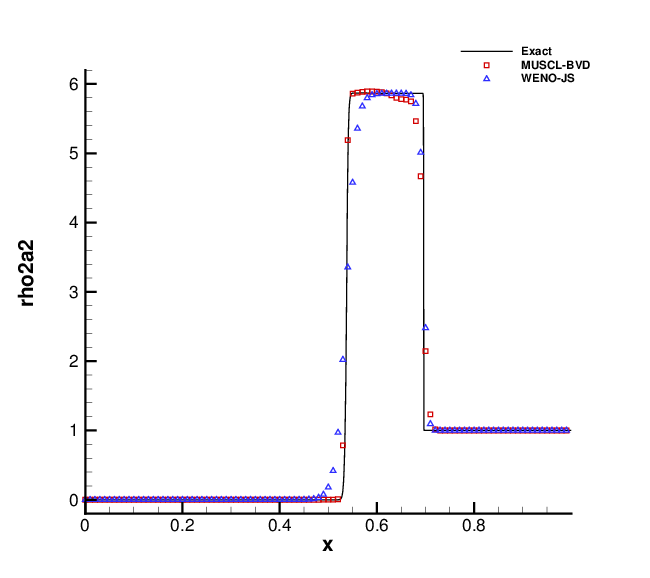}}
	\qquad{}
	\subfigure[$p$]{\centering\includegraphics[scale=0.35]{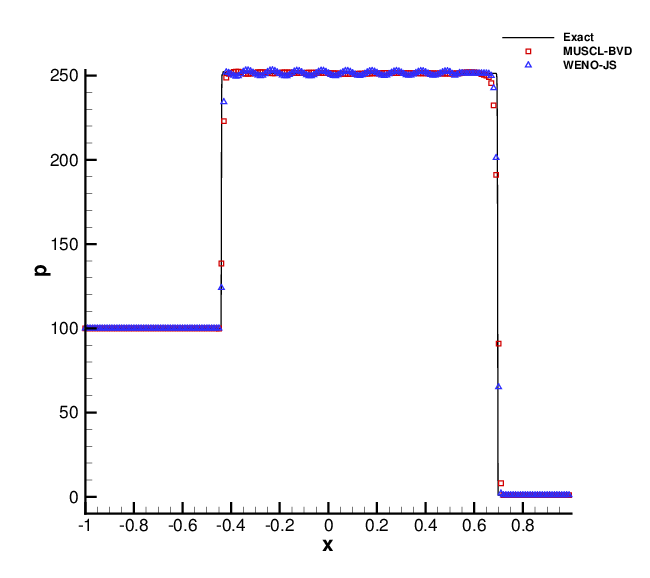}}
	\subfigure[$u$]{\centering\includegraphics[scale=0.35]{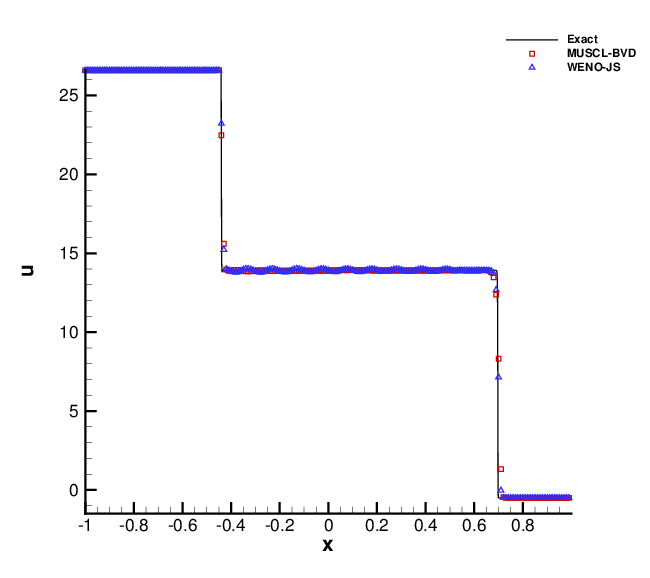}}
	\protect\caption{Comparisons of numerical results of shock/interface interaction problem between MUSCL-THINC-BVD and WENO schemes at t=0.07.\label{fig:interface}}	
\end{figure}

The interaction between a strong shock wave in helium and an air/helium interface has been studied. Typically, such problem is very challenging for some interface tracking methods. For example, the schemes which are not conservative on discrete level may miscalculate the position and speed of the waves resulted from the interaction \cite{test3_1}. The initial problem is set the same as \cite{weno1}, where a Mach 8.96 shock wave is traveling in helium toward a material interface with air which is moving toward the shock wave simultaneously. The detail initial configuration is given by
\begin{equation}
\left(\alpha_{1}\rho_{1},\ \alpha_{2}\rho_{2},\ u_{0},\ p_{0},\ \alpha_{1}\right)=\left\{
\begin{array}{l}
\left(0.386,\ 0,\ 26.59,\ 100,\ 1\right) \ \mathrm{for}\ -1 \leq x < -0.8 \\
\left(0.1,\ 0,\ -0.5,\ 1,\ 1\right) \ \ \ \ \ \ \ \ \ \ \mathrm{for}\ -0.8 \leq x < -0.2 \\
\left(0,\ 1,\ -0.5,\ 1,\ 0\right) \ \ \ \ \ \ \ \ \ \  \ \ \ \mathrm{for}\ -0.2 \leq x < 1
\end{array}
\right..
\end{equation}
The calculation domain is $[-1,1]$ which is divided by 200 uniform mesh cells. The solutions at t=0.07 were computed with the CFL number of 0.1. The comparisons of numerical results between MUSCL-THINC-BVD and WENO schemes are presented in Figure \ref{fig:interface}. The results from MUSCL-THINC-BVD show much superior solution quality in resolving material interface without obvious numerical oscillations, while some oscillations are observed in the pressure and velocity fields by in the results of WENO scheme in the region of the reflected shock wave even although  efforts have been made to implement reconstructions to characteristic variables \cite{weno1}. 

\subsection{Shock-bubble interaction}
In this widely used benchmark test \cite{kokh-lago:anti-diffusiv-jcp2010,
mm:mfluid,shyue:jcp06,so:anti,han,Luo}, we investigate the interactions between a  shock and a bubble which involves a shock wave of Mach 1.22 in air impacting a cylindrical bubble of refrigerant-22 (R22) gas. The experimental results can be referred in \cite{Haas}. A planar rightward-moving Mach $1.22$ shock wave in air impacts a stationary R$22$ gas bubble with radius $r_{0} = 25$mm.  The numerical test, both the air and R$22$ are modeled as perfect gases.
Inside the R$22$ gas bubble, the state variables are 
\[
\left (\rho_{1},  \rho_{2}, u,  v, p,  \alpha_{1} \right ) 
= ( 3.863~\mbox{kg/m}^{3},  1.225~\mbox{kg/m}^{3},
0, 0, 1.01325\times 10^{5}~\mbox{Pa},  1-\varepsilon),
\]
while outside the bubble the corresponding parameters are 
\[
\left (\rho_{1}, \rho_{2}, u, v, p, \alpha_{1} \right ) = 
( 3.863~\mbox{kg/m}^{3}, 1.225~\mbox{kg/m}^{3}, 0,  0, 
1.01325\times 10^{5}~\mbox{Pa}, \varepsilon)  
\]
and
\[
\left (\rho_{1}, \rho_{2}, u, v, p, \alpha_{1} \right ) 
= (3.863~\mbox{kg/m}^{3}, 1.686~\mbox{kg/m}^{3},
113.5~\mbox{m/s}, 0, 1.59 \times 10^{5}~\mbox{Pa}, \varepsilon)
\]    
in the pre- and post- shock regions, respectively, where $\varepsilon=10^{-8}$. The mesh size is $\Delta x = \Delta y =\frac{1}{8}$mm which corresponds to a grid-resolution of 400 cells across the bubble diameter.  Zero-gradient boundary conditions are imposed at the left and right boundaries while symmetric boundaries are imposed at the top and bottom boundaries. Schlieren-type images of density gradient, $|\nabla \rho|$, at different time instants are presented in Figs.\ref{fig:shockBubble}-\ref{fig:shockBubble1}, in which comparisons are made among WENO, MUSCL and MUSCL-THINC-BVD schemes. The MUSCL-THINC-BVD scheme maintains much better the compact thickness of the material interfaces and gives large-scale flow structures similar to the results computed from WENO and MUSCL schemes. Moreover, MUSCL-THINC-BVD scheme is able to reproduce finer flow structures due to largely reduced numerical dissipation.  For example, the instability develops along the interface, which then  rolls up and produces small filaments  as shown in Figure~\ref{fig:shockBubble}. These fine structures in flow and interface tend to be smeared out by numerical schemes with excessive numerical dissipation \cite{so:anti} unless high-resolution meshes are used.  Not only the well-resolved material interface, we can also observe that the reflected shock waves and transmitted shock waves can be captured more clearly compared with the original MUSCL schemes and competitive to  WENO shock-capturing scheme. The resolution quality has been improved remarkably by MUSCL-THINC-BVD scheme to reproduce the complex  flow features which are easily diffused out by conventional shock capturing schemes. 

We make comparisons further with published works which were computed on much finer grid. Shown in Fig.~\ref{fig:SBIca} and \ref{fig:SBIcl} we plot our results on a coarse mesh where the diameter uses 400 cells to compare with the results computed by anti-diffusion interface sharpening technique \cite{so:anti} and multi-scale sharp interface \cite{Luo} on a finer grid where 1150 cell were used for the bubble diameter. From Fig. \ref{fig:SBIcl}, it can be observed that similar small-scale structures have been recovered by the MUSCl-THINC-BVD scheme with much fewer cells.  

\begin{figure}
		\subfigure[t=210 $\mu$s]{\centering\includegraphics[scale=0.14]{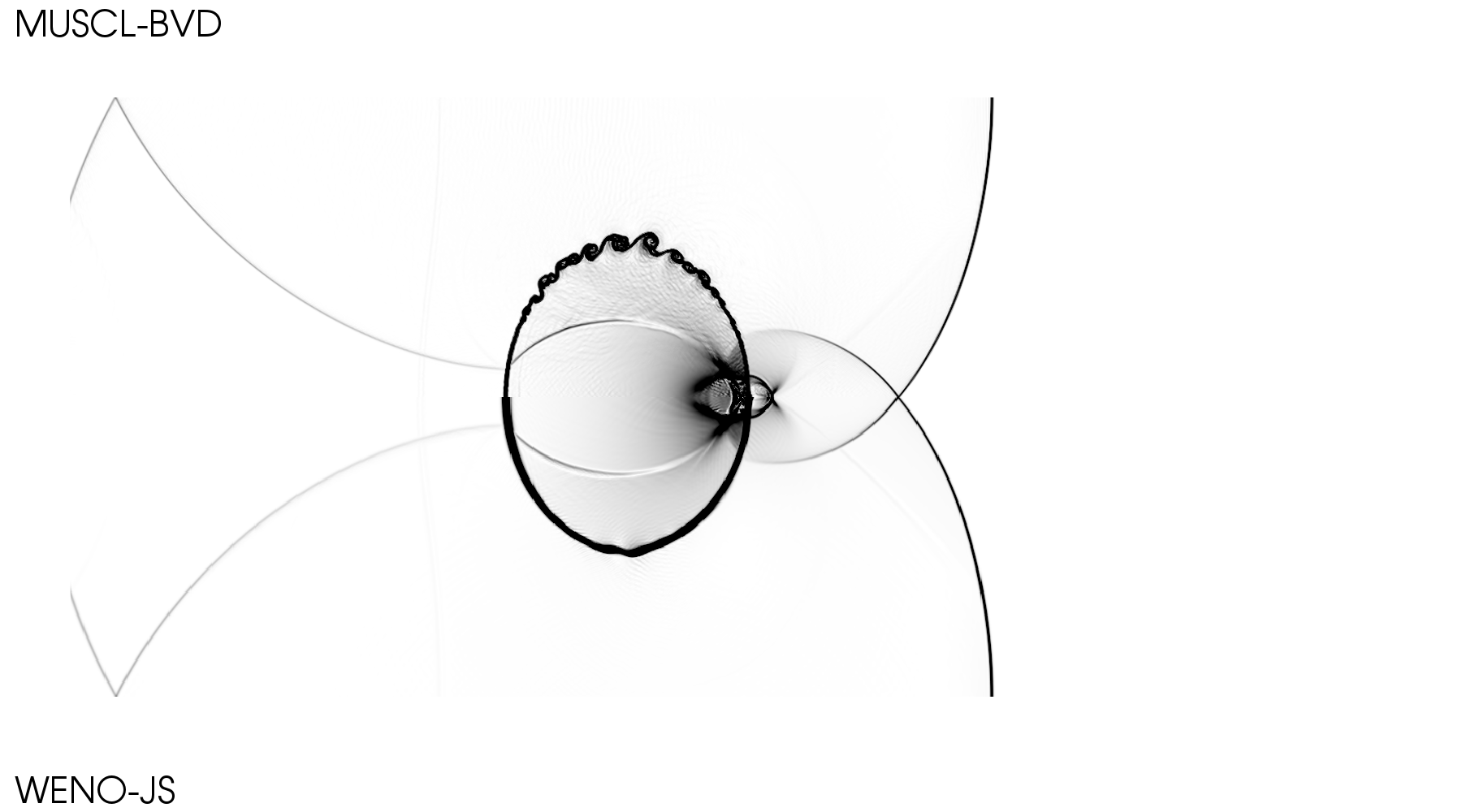}}
		\subfigure[t=210 $\mu$s]{\centering\includegraphics[scale=0.14]{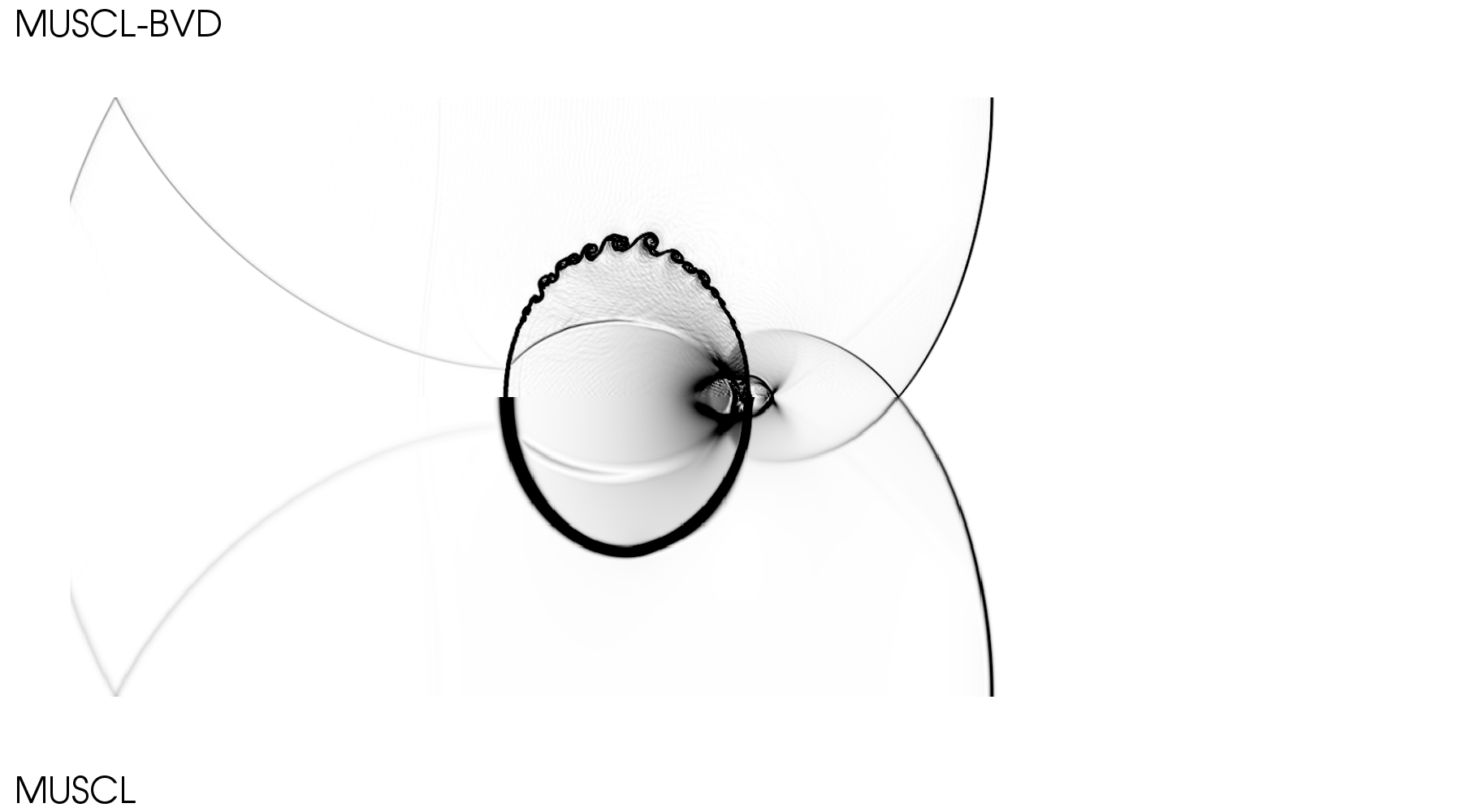}}
		    \qquad{}
		\subfigure[t=264 $\mu$s]{\centering\includegraphics[scale=0.14]{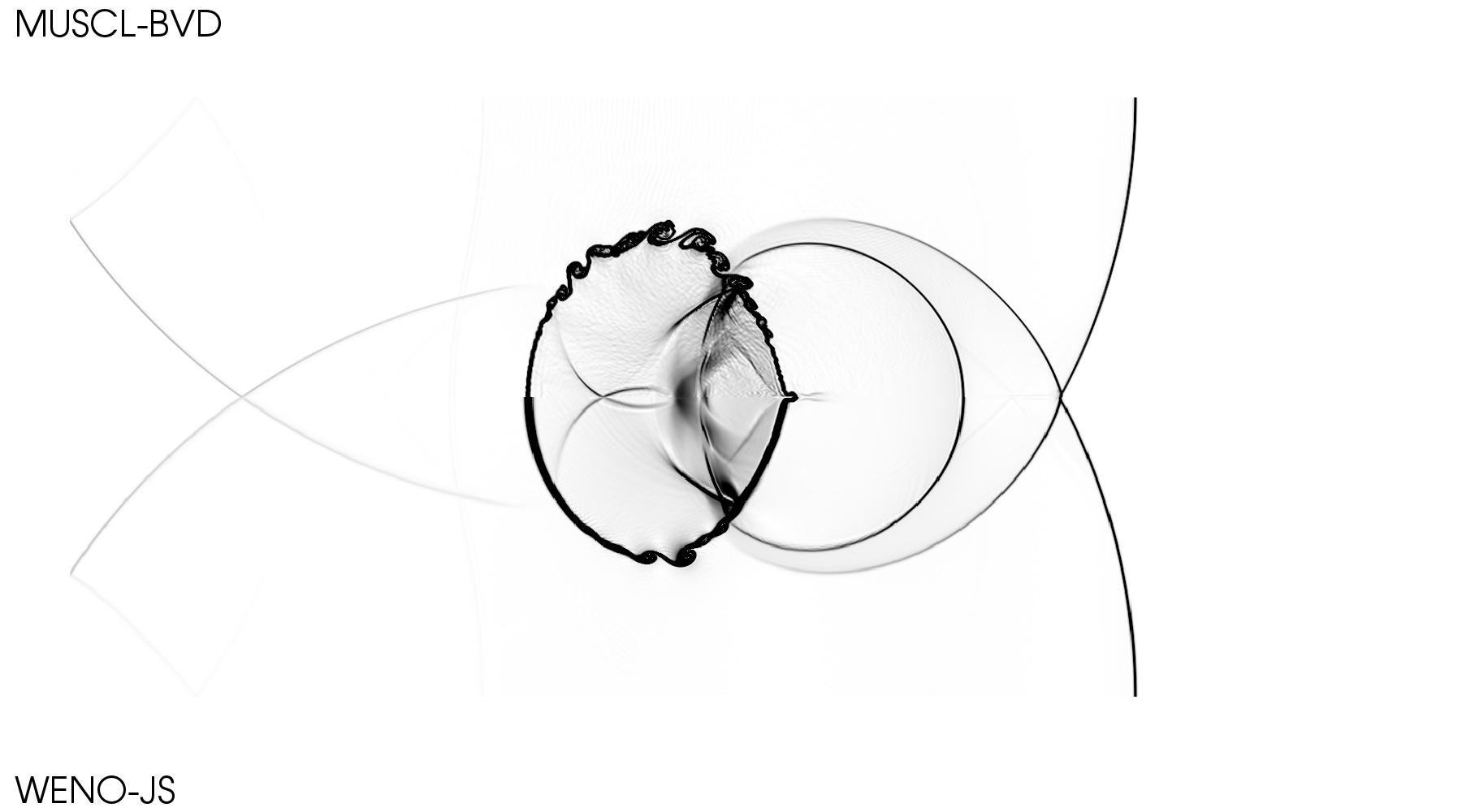}}
\subfigure[t=264 $\mu$s]{\centering\includegraphics[scale=0.14]{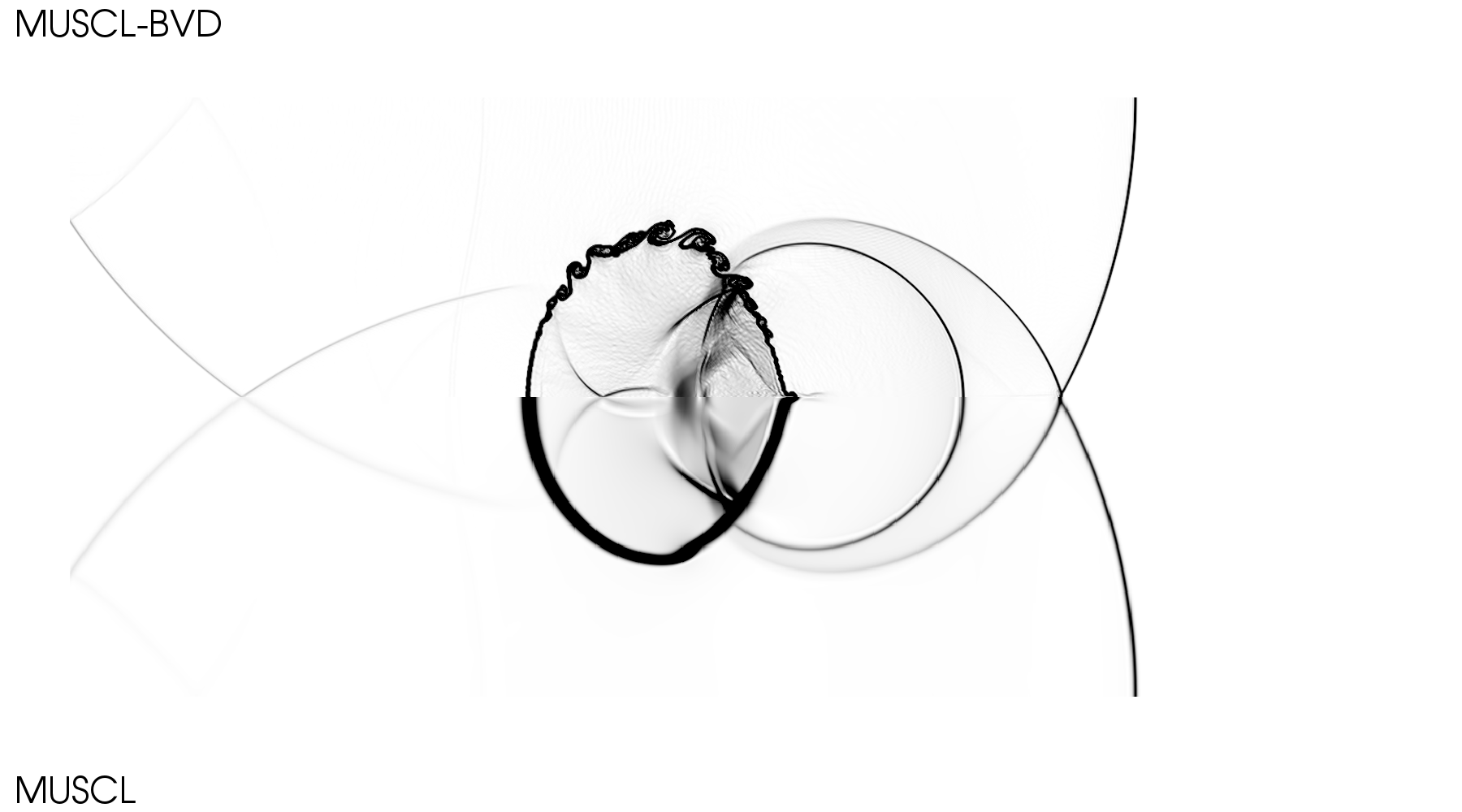}}
\qquad{}
		\subfigure[t=296 $\mu$s]{\centering\includegraphics[scale=0.14]{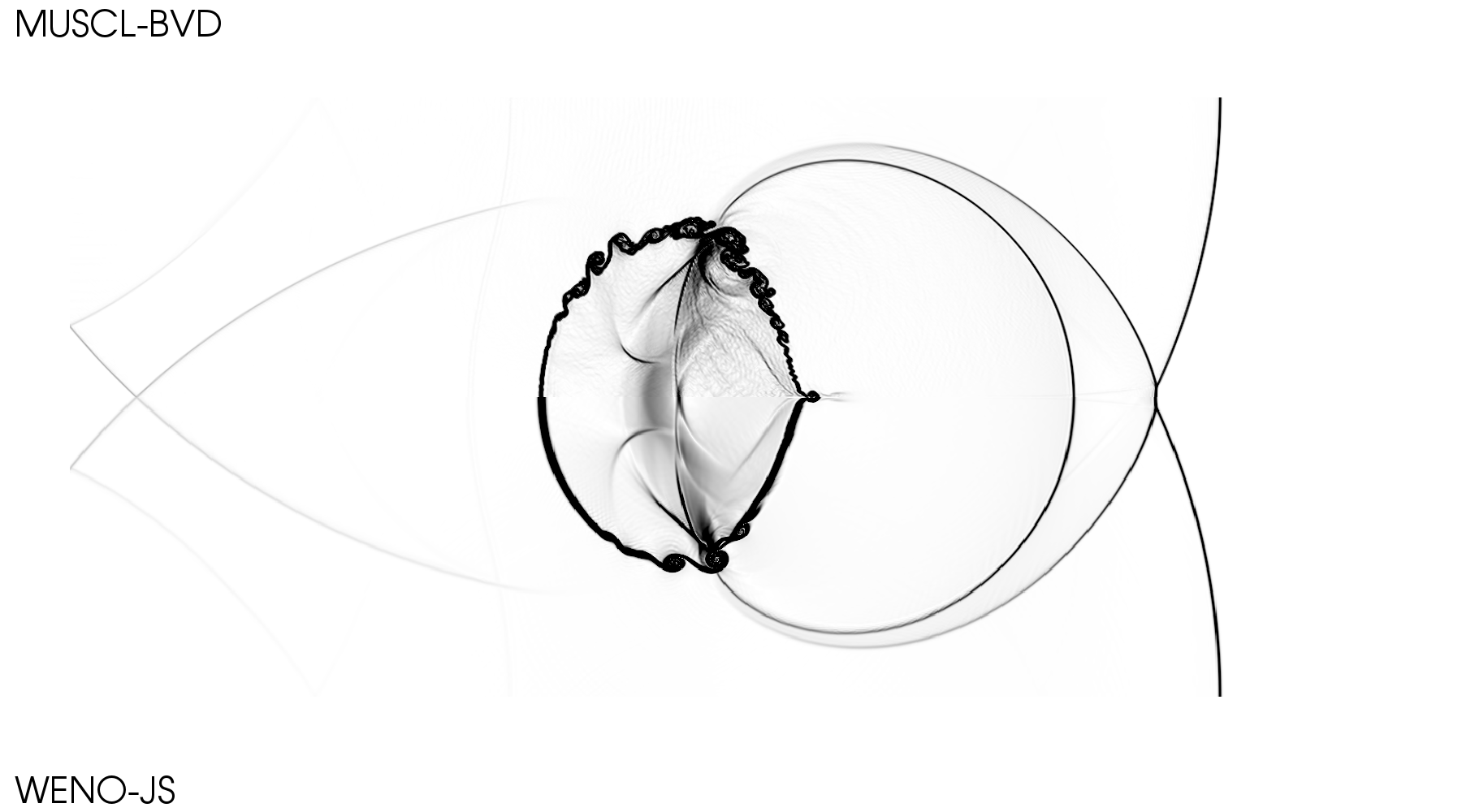}}
      \subfigure[t=296 $\mu$s]{\centering\includegraphics[scale=0.14]{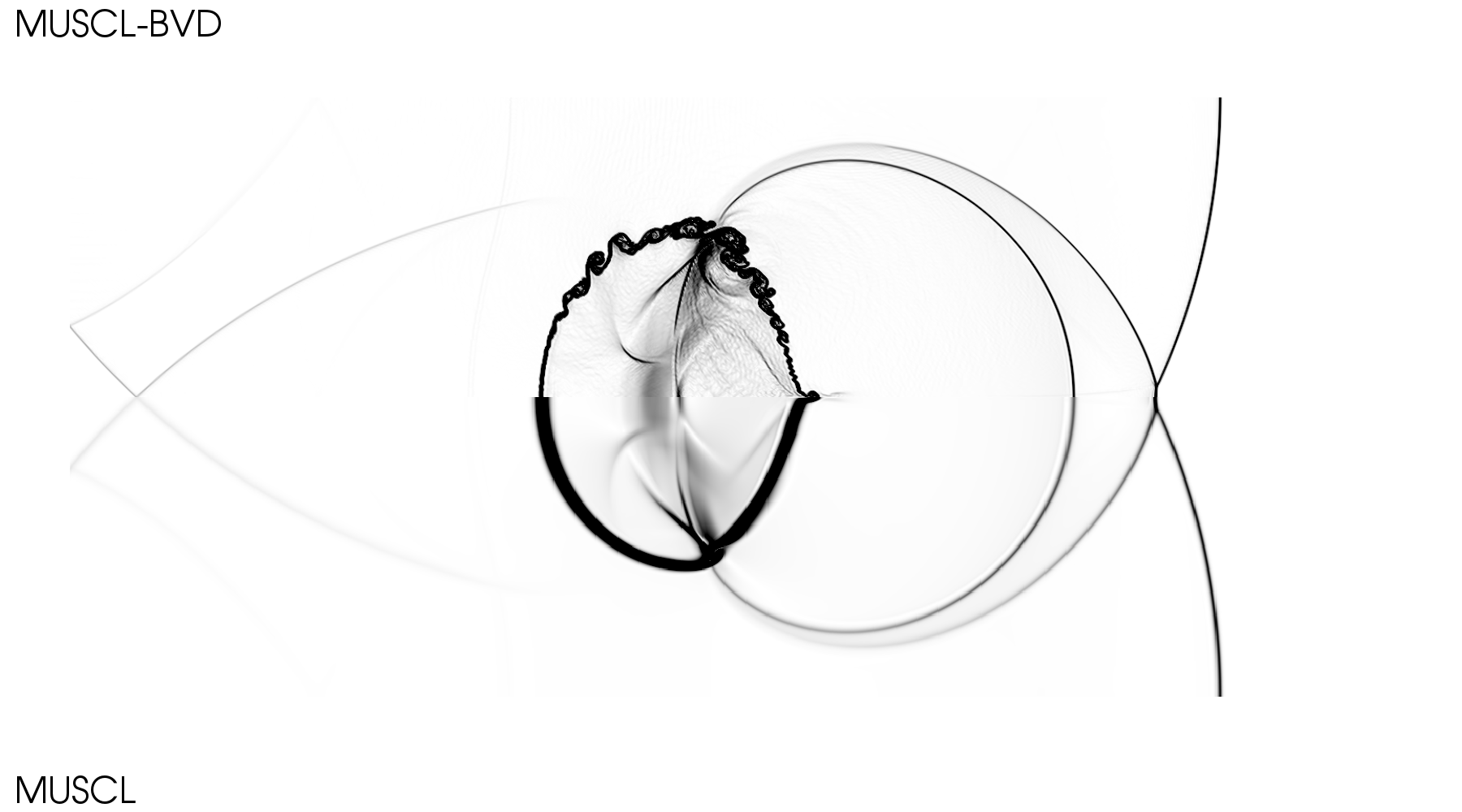}}
		   \protect\caption{Numerical results for a planar Mach $1.22$ shock wave in air interacting with a circular R22 gas bubble. Comparisons are made among MUSCL-THINC-BVD, WENO-JS and MUSCL schemes regarding to Schlieren-type images of density at different times\label{fig:shockBubble}}	
\end{figure}

\begin{figure}
	\subfigure[t=372 $\mu$s]{\centering\includegraphics[scale=0.14]{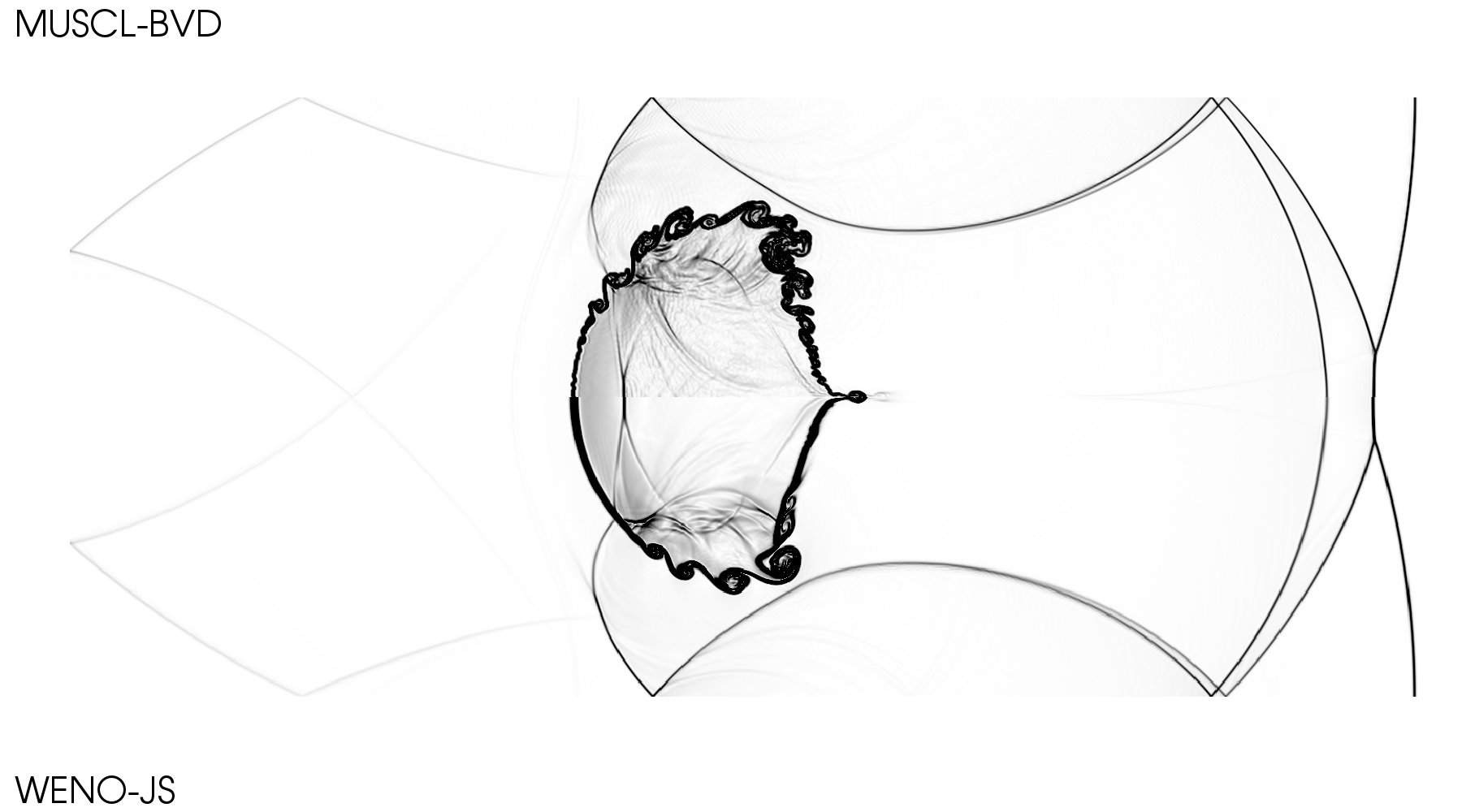}}
	\subfigure[t=372 $\mu$s]{\centering\includegraphics[scale=0.14]{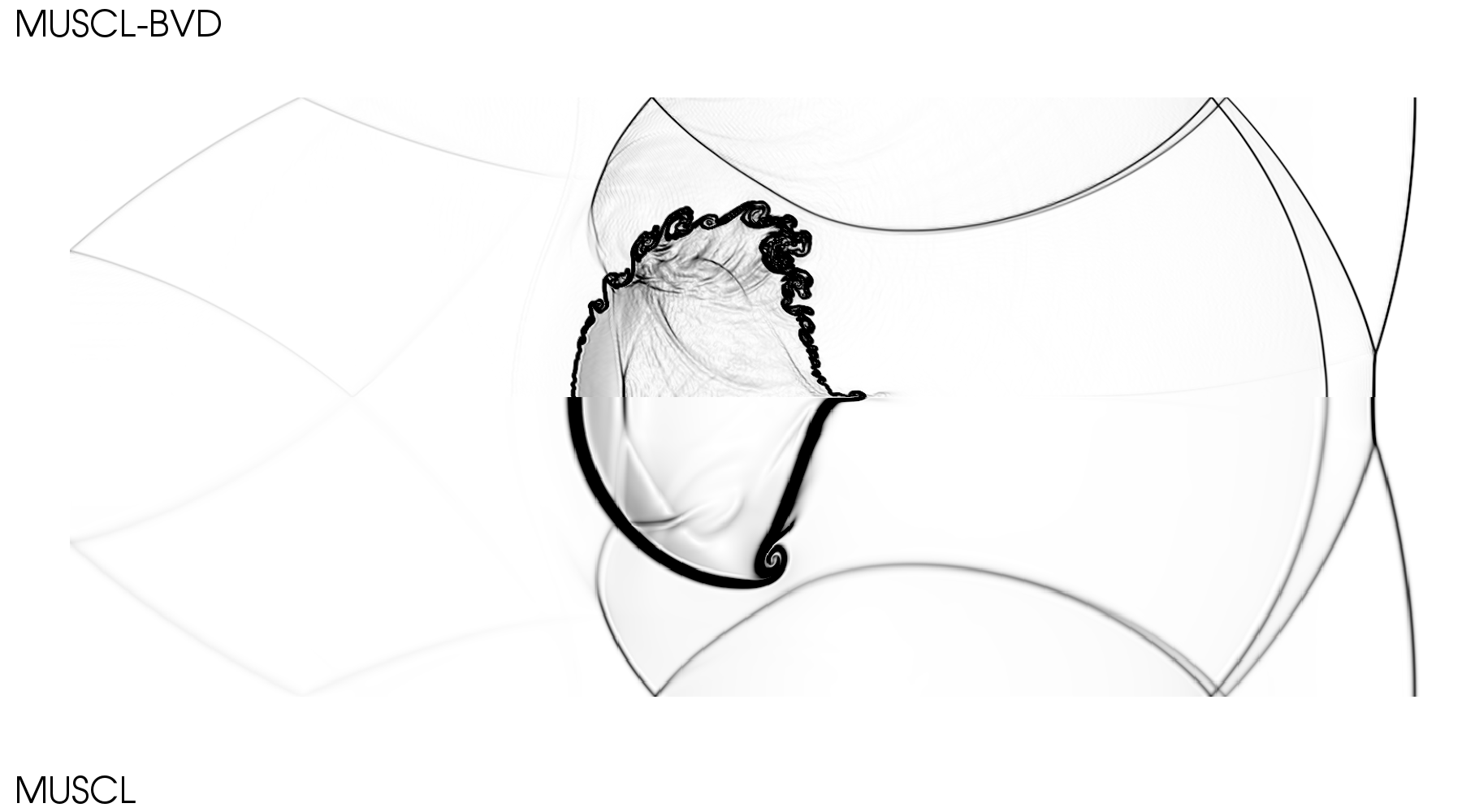}}
	\protect\caption{(continued)\label{fig:shockBubble1}}
		
\end{figure}

\begin{figure}
	\subfigure[t=187 $\mu$s]{\centering\includegraphics[scale=0.5]{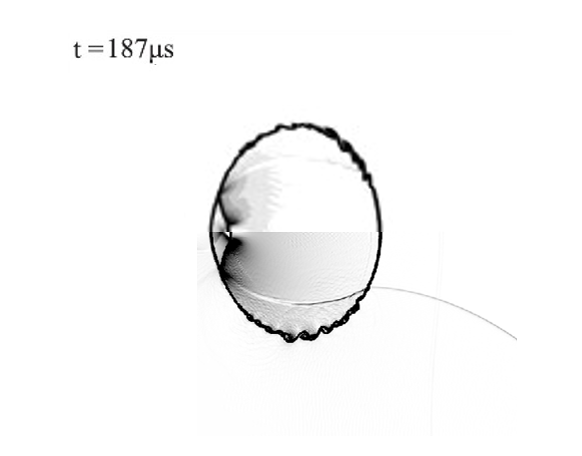}}
	\subfigure[t=247 $\mu$s]{\centering\includegraphics[scale=0.5]{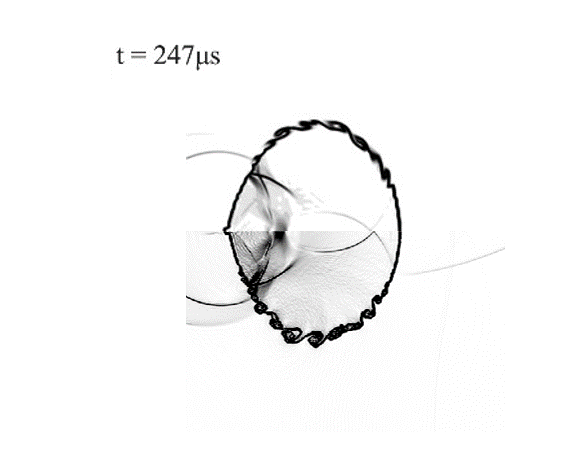}}
	\qquad{}
	\subfigure[t=318 $\mu$s]{\centering\includegraphics[scale=0.5]{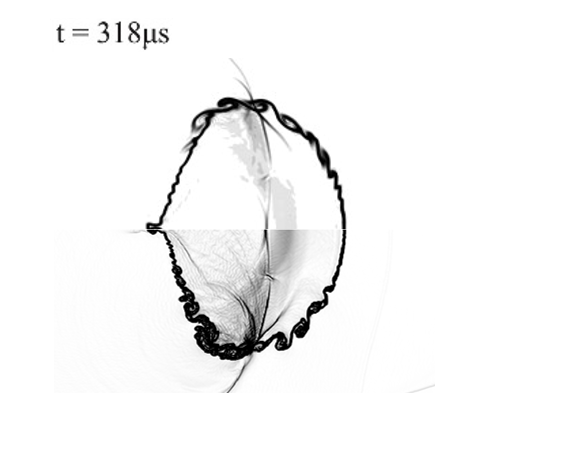}}
	\subfigure[t=417 $\mu$s]{\centering\includegraphics[scale=0.5]{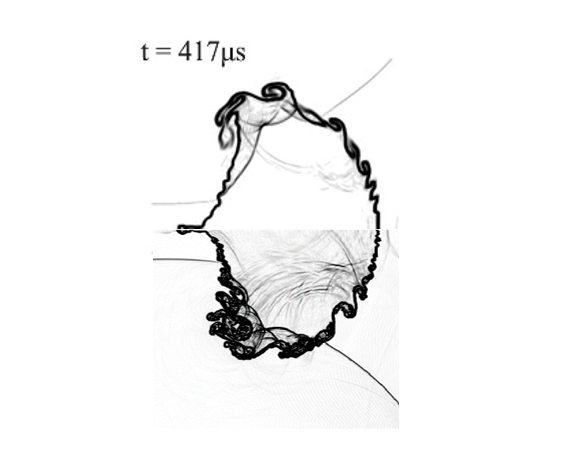}}
	\protect\caption{Comparisons with published work \cite{so:anti} about the Schlieren-type images with the same grid number. The top half is the result calculated by anti-diffusion interface sharpening technique while the bottom half is from the MUSCL-THINC-BVD scheme  \label{fig:SBIca}}	
\end{figure}

\begin{figure}
	\subfigure[t=187 $\mu$s]{\centering\includegraphics[scale=0.5]{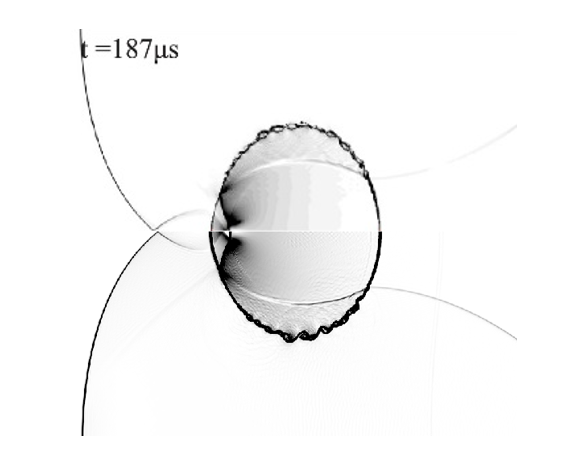}}
	\subfigure[t=247 $\mu$s]{\centering\includegraphics[scale=0.5]{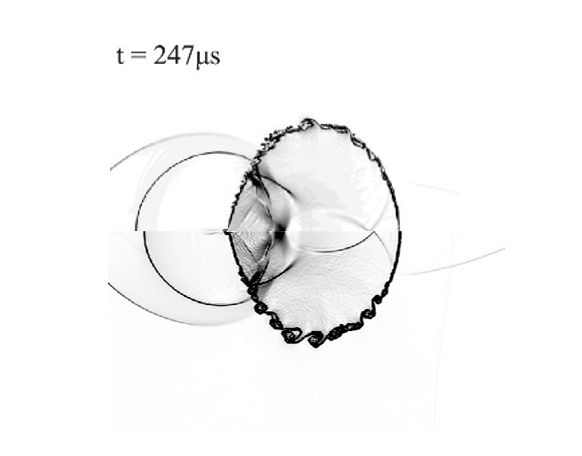}}
	\qquad{}
	\subfigure[t=318 $\mu$s]{\centering\includegraphics[scale=0.5]{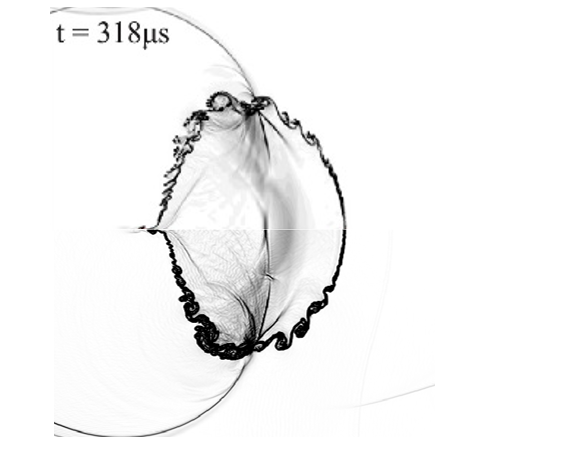}}
	\subfigure[t=417 $\mu$s]{\centering\includegraphics[scale=0.5]{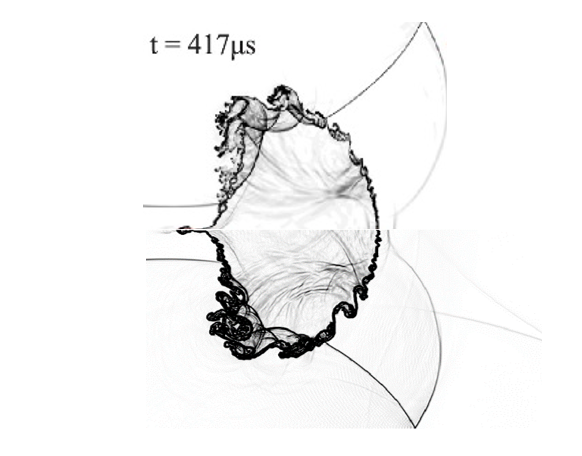}}
	\protect\caption{Comparisons with published work \cite{Luo} about the Schlieren-type images. The top half is the result calculated by multi-scale sharp interface modeling with 1150 cells distributed along the bubble diameter while the bottom half is from the MUSCL-THINC-BVD scheme with 400 cells along the bubble diameter.\label{fig:SBIcl}}	
\end{figure}

\section{Conclusion remarks \label{sec:conclusion}}
In this work, we implement MUSCL-THINC-BVD scheme to simulate compressible multiphase flows by solving the five-equation model. This scheme can resolve discontinuous solutions with much less numerical dissipation. By treating interface as another contact discontinuity rather than implementing interface-sharping techniques explicitly, the new scheme can realize thermodynamical-consistent reconstruction straightforwardly. The results of test cases show a remarkable improvement in the solution quality to the problems of interest. Compared with the high-order shock-capturing schemes, the new scheme shows competitive or even better numerical results but with less computational cost. This work provides an effective but simple approach to simulate compressible interfacial multiphase flows.

\section*{Acknowledgment}

This work was supported in part by JSPS KAKENHI Grant Numbers 15H03916 and 15J09915. 

\clearpage{}


\end{document}